\documentclass[preprint, showpacs, prb]{revtex4-1}
\usepackage{bbm}
\usepackage{amsmath}
\usepackage{amssymb}
\usepackage{graphicx}
\usepackage{multirow}
\usepackage{hyperref}

\begin{document}
\title{Enhancement of superconductivity by external magnetic field in magnetic alloys}
\date{\today}
\author{Dawid Borycki}
\email{dawid.borycki@fizyka.umk.pl}
\address{Instytut Fizyki, Uniwersytet M. Kopernika, ul. Grudziadzka 5, 87-100 Torun, Poland}
\date{\today}


\begin{abstract}
An asymptotic solution of the thermodynamics of a BCS superconductor containing spin $1/2$ and $7/2$ magnetic impurities, obtained recently in Ref.~\onlinecite{DB11} is exploited to derive the expressions for critical magnetic field $H_{\mathrm{c}}(T)$ . The credibility of the resulting theoretical equations, which depend on the magnetic coupling constant $g$ and impurity concentration $c$, is verified on the experimental data for the following superconducting alloys: LaCe, ThGd and SmRh$_{4}$B$_{4}$. Good quantitative agreement with experimental data is found for sufficiently small values of $c$. The discrepancies between theoretical and experimental values of $H_{\mathrm{c}}(T)$ for larger values of $c$ in case of LaCe and ThGd are reduced by introducing the concept of the effective temperature $\tilde{T}$, which accounts for the Coulomb interactions between the electron gas and impurity ions.

At low temperatures, the critical magnetic field is found to increase with decreasing temperature $T$. This enhancement of the critical magnetic field provides evidence of the Jaccarino-Peter effect, which was experimentally observed in the Kondo systems like LaCe, (La$_{1-x}$Ce$_{x}$)Al$_{2}$ and also in the pseudoternary compounds, including Sn$_{1-x}$Eu$_{x}$Mo$_{6}$S$_{8}$, Pb$_{1-x}$Eu$_{x}$Mo$_{6}$S$_{8}$ and La$_{1.2-x}$Eu$_{x}$Mo$_{6}$S$_{8}$.

The effect of an external magnetic field ${\mathcal H}$ on a BCS superconductor perturbed by magnetic impurities was also studied. On this grounds, by analyzing the dependence of superconducting transition temperature $T_{\mathrm{c}}$ on ${\mathcal H}$ of (La$_{1-x}$Ce$_{x}$)Al$_{2}$, we have shown, that for certain parameter values, external magnetic field compensates the destructive effect of magnetic impurities.
\end{abstract}

\maketitle

\section{Introduction}
The effect of superconductivity has been discovered over 100 years ago \cite{HK11}. Since then, many extraordinary properties accompanying this phenomenon, including perfect diamagnetism, zero-resistance, magnetic levitation due to expulsion of the magnetic field from a superconductor, flux quantization, Josephson effect and vortex state have been explained.

The promising future applications of superconductors mainly concern electric power applications \cite{DL01} and generation of high, uniform magnetic field, e.g. for magnetic resonance imaging purposes~\cite{HM84W}. However, the wide-scale applications of superconductors are limited, because superconductivity has been proven to be very sensitive to the destructive effect of an external magnetic field and the current density. In order to extend the applications of superconductors it is strongly desirable to enhance the values of the critical magnetic field, critical currents and superconducting transition temperature. These topics were of high interest during past few decades\cite{Matthias1, Matthias2, Matthias3, OF82, BM82, BM, Hosono, HR03, MapleConvAndUnconvSC, Chen2010, Gardner2011, DA011}.

The superconducting critical temperature, as shown by Bednorz and M\"{u}ller \cite{BM} and later by Hosono group \cite{Hosono}, can be raised by adding new elements to the antiferromagnetically ordered host system. Superconductivity appears at small value of dopant concentration $x$. Subsequently, the transition temperature $T_{\mathrm{c}}$ increases almost linearly with $x$, and after reaching a maximum at optimal doping level $x_{opt}$, decreases and finally falls to zero. According to this scenario, the existence of a parent magnetic correlations is viewed to be an essential feature of high-temerature superconductivity.

On the other hand, Matthias et al.~\cite{Matthias1} discovered that the superconducting transition temperature of lanthanum decreases, when small amount (1 at. \%) of the rare-earth magnetic impurities are added. It was shown, that the depression of $T_{\mathrm{c}}$ increases with the impurity spin value and not, as it was expected, with the impurity magnetic moment.

The experimental studies carried out by Matthias revealed another extraordinary property of superconducting alloys, namely the possible coexistence of magnetism and superconductivity. The coexistence of superconductivity and long-range antiferromagnetic ordering of the rare earth R magnetic moments was later discovered in RMo$_{6}$Se$_{8}$ (R = Gd, Tb and Er)~\cite{McCallumASC}, RRh$_{4}$B$_{4}$ (R = Nd, Sm and Tm)~\cite{HamakerASC} and in RMo$_{6}$S$_{8}$ (R = Gd, Tb, Dy and Er)~\cite{IshikawaASC}. A~similar overlap between superconductivity and ferromagnetism was observed in ErRh$_{4}$B$_{4}$~\cite{MapleFSC} and HoMo$_{6}$S$_{8}$~\cite{IshikawaFSC}.

This discovery was surprising, since magnetism and superconductivity had been believed to be mutually exclusive, because the internal magnetic fields generated in magnetically ordered systems are much larger than the typical critical fields of superconductors. However, as independently predicted by Fulde and Ferrell~\cite{FF64} and Larkin and Ovchinnikov~\cite{LO65}, the superconductor may overcome the pair-breaking effect of magnetic field by forming periodic regions of superconductivity separated by domains of aligned spins. In such FFLO-state the order parameter is spatially modulated along the field direction. There are strongly experimental suggestions for the occurrence of FFLO state in some heavy-fermion compounds, e.g.~in~CeCoIn$_{5}$~\cite{AB02}.

Paradoxically, under specific circumstances, the external magnetic field can even enhance (instead of depress) the properties of superconductors, e.g. the upper critical field $H_{\mathrm{c}2}$ of SmRh$_{4}$B$_{4}$ increases below the N\'{e}el temperature $T_{\mathrm{N}} = 0.87$ K~\cite{OF82, BM82}. The latter can be understood on the grounds of the Jaccarino-Peter effect \cite{VJ62}, in which the external magnetic field compensates the antiferromagnetic exchange interaction generated by the conduction electrons antiferromagnetically coupled to the spins of localized magnetic moments. The Jaccarino-peter compensation effect has been observed in several systems, e.g. in Eu$_{0.75}$Sn$_{0.25}$Mo$_{6}$S$_{7.2}$Se$_{0.8}$~\cite{HM84}.

In this study, we shall not investigate the FFLO state, since we assume a spatially uniform superconducting order parameter. In such the case, only the Jaccarino-Peter effect can be considered as the possible source of the superconductivity enhancement. Before going into details, let us briefly summarize theoretical investigations of the superconductors containing magnetic impurities.

Early theoretical investigations of the problem of magnetic superconducting alloys were founded on perturbation theory. Nakamura~\cite{Nakamura} and Suhl et al.~\cite{Suhl} explained this effect by treating the s-d interaction $V_{\mathrm{s-d}}$~\cite{Kasuya} as an additive term in the total Hamiltonian, which perturbs a BCS superconductor~\cite{BCS}. Balseiro et al. \cite{Balseiro} studied a BCS superconductor perburbed by magnetic impurities interacting via a~nearest neighbour Heisenberg potential. The resulting phase diagrams comply qualitatively with experiment.

The well-known Abrikosov-Gor'kov theory~\cite{AG} (AG) of dirty superconductors explains the strong decrease in $T_{\mathrm{c}}$ due to magnetic impurities and also predicts {\itshape gapless superconductivity}, confirmed experimentally by Reif and Woolf~\cite{ReifWoolf}. Disagreement with this approach is observed in a number of Kondo superconductors, e.g. La$_{1-x}$Ce$_{x}$Al$_{2}$~\cite{MapleConvAndUnconvSC}, LaCe and LaGd~\cite{ChaikinMihalisin} and PbCe and InCe~\cite{DelfsBeaudryFinnemore}.

The AG theory was generalized to describe an increase of $H_{\mathrm{c}2}$ below the N\'{e}el temperature $T_{\mathrm{N}}$. To this end, Ramakrishnan and Varma~\cite{RV81} extended AG theory to the case, when the magnetic ions are present in large concentration. They solved numerically Eliashberg equations including the effects of phonons, spin waves and elastic scattering in order to estimate the variations of the pairbreaking parameter. However, a detailed comparison of their results with experimental data for the upper critical field of SmRh$_{4}$B$_{4}$ have been not performed.

Hamaker et al. \cite{HWMFM78} found a good fit of the expression for the upper critical magnetic field in Machida's theory for antiferromagnetic superconductors to their experimental data on SmRh$_{4}$B$_{4}$ and obtained good quantitative agreement. However, some of the adjustable parameter values does not comply with other experimental findings.



The effect of magnetic impurities on superconductivity is still under debate. Kozorezov et al.~\cite{AK08} have shown, that trace concentrations of magnetic impurities may also result in significant changes in nonequilibrium properties of superconductors. A~comprehensive review of recent developments in this field can be found in Ref.~\onlinecite{AB06}.


These yet unresolved issues, as well as some shortcomings of the models presented above, motivate the present work. We investigate the thermodynamics of the Hamiltonian
\begin{equation}
H^{(M)} = H_{\mathrm{BCS}} + V^{(M)},
\end{equation}
where
\begin{equation}
H_{\mathrm{BCS}} = H_{0} + V_{\mathrm{BCS}},
\end{equation}
and
\begin{equation}
H_{0} = \sum_{\mathbf{k}\sigma} \xi_{\mathbf{k}} n_{\mathbf{k}\sigma},
\end{equation}
with $\xi_{\mathbf{k}} = \varepsilon_{\mathbf{k}} - \mu$, $n_{\mathbf{k}\sigma} = a_{\mathbf{k}\sigma}^{\dagger}a_{\mathbf{k}\sigma}^{}$, is the free fermion kinetic energy operator and
\begin{equation}
\label{VBCS}
V_{\mathrm{BCS}}^{(M)} = - |\Lambda|^{-1} \sum_{\mathbf{k} \mathbf{k^{\prime}}} G_{\mathbf{k}\, \mathbf{k^{\prime}}} a_{\mathbf{k}+}^{\dagger} a_{-\mathbf{k}-}^{\dagger} a_{-\mathbf{k}^{\prime}-}^{} a_{\mathbf{k}^{\prime}+},
\end{equation}
is the Cooper pairing potential, whereas
\begin{equation}
\label{SDreduced}
V = g^{2} N^{-1} \sum_{\alpha=1}^{M} \sigma_{z} S_{z\alpha},
\end{equation}
represents the reduced s-d interaction. $|\Lambda|$ denotes the system's volume and $G_{\mathbf{k}\, \mathbf{k}'}^{}$ is real, symmetric, invariant under $\mathbf{k}\rightarrow - \mathbf{k}$ or $\mathbf{k}'\rightarrow - \mathbf{k}'$ and nonvanishing only in a thin band close to the Fermi surface, viz.,
\[
G_{\mathbf{k}\, \mathbf{k}'}^{} = G_{0}\chi(\mathbf{k})\chi(\mathbf{k}'), \qquad G_{0} > 0,
\]
where $\chi(\mathbf{k})$ denotes the characteristic function of the set
\[
{\cal P} = \{\mathbf{k}: \mu - \delta \leq \varepsilon_{\mathbf{k}}^{} \leq \mu + \delta \}, \qquad \varepsilon_{\mathbf{k}}^{} = \frac{\hbar^{2} \mathbf{k}^{2}}{2m}.
\]
In equation (\ref{SDreduced}), $S_{z\alpha}$ denotes the spin operator of the $\alpha-$th magnetic ion, whereas
\[
\sigma_{z} = \sum_{\mathbf{k}\in{\cal P}}\left(n_{\mathbf{k+}} - n_{\mathbf{k-}}\right),
\]
is the spin operator of a conducting fermion. $M$ is the number of magnetic impurities, $N$ the number of host atoms.

We assume the perturbation implemented by the localized distinguishable magnetic impurities to be a reduced long-range s-d interaction, which involves only the z-components of the impurity and fermion spin operators (Eq.~(\ref{SDreduced})). The reason for this simplification is that the thermodynamics of the resulting Hamiltonian $H^{(M)}=H_{0} + V_{\mathrm{BCS}} + V^{(M)}$ admits a mean-field solution, the accuracy of which improves with decreasing impurity density. Furthermore, this solution is thermodynamically equivalent to the one obtained for $H$ with a Heisenberg type reduced s-d interaction
\begin{equation}
V_{\mathrm{H}}^{} = -\frac{g^2}{N} \sum_{\mathbf{k}\in{\cal P},\alpha} \Bigl[(a_{\mathbf{k}-}^{\dagger} a_{\mathbf{k}-}^{} - a_{\mathbf{k}+}^{\dagger} a_{\mathbf{k}+}^{} S_{z\alpha}^{} -  a_{\mathbf{k}+}^{\dagger} a_{\mathbf{k}-}^{}S_{\alpha -}^{} - a_{\mathbf{k}-}^{\dagger} a_{\mathbf{k}+}^{}S_{\alpha +}^{}\Bigr],
\end{equation}
\[
S_{\alpha \pm}^{} = S_{\alpha x}^{} \pm \mathrm{i} S_{\alpha y}^{},
\]
replacing $V$ (Ref.~\onlinecite{JM99}, Sec. 6.2.5). (Similarly, the thermodynamics of a classical superconductor can be explained in terms of a reduced BCS interaction, whereas a~gauge-invariant theory of the Meissner effect requires a more general pairing potential.) The reduced form of $V_{H}$, obtained by rejecting the sum $\sum\limits_{\mathbf{k} \neq \mathbf{k}^{\prime}} V_{\mathbf{k} \, \mathbf{k}^{\prime}}$ in the s-d interaction $V_{\mathrm{s-d}}$ and restricting the remaining sum $\sum\limits_{\mathbf{k}} V_{\mathbf{k} \, \mathbf{k}}$ to a thin layer, is an approximation which resorts to the fact that spin-exchange processes, and not momentum exchange processes, are primarily responsible for the Kondo effect caused by $V_{\mathrm{s-d}}$ \cite{Kondo} and to the low-temperature regime applied.

In section~\ref{Sec:UpperBound} the system's free energy $F(H, \beta)$ is determined within the Bogolyubov inequality. The resulting approximating formula for the free energy is used in sections~\ref{Sec:MeanFieldDescription} and \ref{Sec:CriticalMagneticField} to derive the expression for the critical magnetic field $H_{\mathrm{c}}$. The credibility of the theoretical expression for $H_{\mathrm{c}}$, which depends on the magnetic coupling constant $g$ and impurity concentration $c$, is subsequently verified on the experimental data for LaCe, ThGd and SmRh$_{4}$B$_{4}$. Good quantitative agreement with experiment has been obtained. At low temperatures, the critical magnetic field is found to increase with decreasing temperature, similarly as in some antiferromagnetic superconductors.

In section~\ref{Sec:SCH} the effect of an external magnetic field on superconducting alloys is studied. The resulting expression for the superconducting transition temperature $T_{\mathrm{c}}({\mathcal H})$ is exploited in section~\ref{Sec:CriticalTemperature} in order to explain the compensation effect.

The theory presented improves earlier developments in this field. Apart from an explanation of the reentrant behavior of superconducting alloys~\cite{DB11}, it provides an explanation of the Jaccarino-Peter effect and good quantitative agreement with the experimentally measured critical magnetic field of several superconducting alloys LaCe, ThGd and SmRh$_{4}$B$_{4}$.

\section{Upper bound to the free energy in terms of the Bogolyubov method}
\label{Sec:UpperBound}
The full Hamiltonian of the system
\begin{equation}
\label{Hsd}
H^{(M)} = H_{0} + V_{\mathrm{BCS}} + V^{(M)}
\end{equation}
can be expressed in the following form in terms of mean-field parameters $\nu$, $\eta$:
\begin{equation}
\label{H_M}
H^{(M)} = h^{(M)}(\nu, \eta) + H_{R}^{(M)},
\end{equation}
where
\begin{equation}
\label{hM}
h^{(M)}(\nu, \eta) = \tilde{h} + h_{imp}^{(M)} + \frac{1}{2}MN(\nu^{2} - \eta^{2}),
\end{equation}
\begin{equation}
\label{hEl}
\tilde{h} = H_{\mathrm{BCS}} + \kappa \sigma_{z}, \qquad \kappa = - g M (\nu - \eta)
\end{equation}
\begin{equation}
\label{hImp}
h_{imp}^{(M)} = g \nu \sum_{\alpha}S_{z\alpha} + \frac{1}{2}N^{-1}g^{2}\sum_{\alpha}S_{z\alpha}^{2},
\end{equation}
\begin{equation}
\label{HR_M}
H_{R}^{(M)} = - \frac{1}{2}N^{-1} \sum_{\alpha} \Bigl[g\bigl(\sigma_{z} - S_{z\alpha}\bigr) - \nu N\Bigr]^{2} + \frac{1}{2} N^{-1} \sum_{\alpha}\bigl(g \sigma_{z} - \eta N\bigr)^{2}.
\end{equation}
\noindent The Bogolyubov inequality
\begin{equation}
\label{Bogolubow}
F(H_{1}+H_{2}) \leq F(H_{1}) + \left<H_{2}\right>_{H_{1}},
\end{equation}
with $H_{1} = h^{(M)}(\nu, \eta)$, yields
\begin{equation}
\label{FF}
F(H^{(M)}, \beta) \leq  F(h^{(M)}(\nu, \eta), \beta) + \left<H^{(R)}\right>_{h^{(M)}}.
\end{equation}
The parameters $\nu$ and $\eta$ will be now choosen so that they minimize the free energy $F(h^{(M)}(\nu, \eta), \beta) = -\beta^{-1} \ln \text{Tr} \exp \left[-\beta h^{(M)}(\nu, \eta) \right]$, viz.,
\[
\frac{\partial F(h^{(M)}(\nu, \eta), \beta)}{\partial \nu} = 0, \qquad \frac{\partial F(h^{(M)}(\nu, \eta), \beta)}{\partial \eta} = 0.
\]
The explicit form of these equations is
\begin{equation}
\label{nu}
\nu = \frac{g}{N} \left<\sigma_{z}\right>_{\tilde{h}} - \frac{g}{N} \frac{\text{Tr} S_{z} \exp\left[-\beta h_{imp}\right]}{\text{Tr} \exp[-\beta h_{imp}]},
\end{equation}
\begin{equation}
\label{eta}
\eta = \frac{g}{N} \left<\sigma_{z}\right>_{\tilde{h}}.
\end{equation}
Using equations (\ref{HR_M}), (\ref{nu}) and (\ref{eta}) one obtains
\begin{equation}
\label{Hr}
\left<H^{(R)}\right>_{h^{(M)}} = -\frac{1}{2} c g^{2} \left(\left<S_{z}^{2}\right>_{h_{imp}^{(1)}} - \left<S_{z}\right>^{2}_{h_{imp}^{(1)}}\right).
\end{equation}
The inequality $\text{Tr}(\rho A^{2}) \geq \left(\text{Tr}(\rho A)\right)^{2},$ valid for any bounded self-adjoint operator $A$ and density matrix $\rho$, shows that $\left<H^{(R)}\right>_{h^{(M)}} \leq 0$. Hence, from equations~(\ref{FF}),~(\ref{Hr}) one obtains the relevant upper bound to the free energy
\begin{equation}
\label{UpperBound}
F(H^{(M)}, \beta) \leq  F(h^{(M)}(\nu, \eta), \beta).
\end{equation}
According to Eq.~(\ref{UpperBound}) we ascertain that the thermodynamics of the original system, characterized by $H^{(M)}$, is almost equivalent to that of $h^{(M)}$, provided $\eta$ and $\nu$ are the minimizing solutions of the equations (\ref{nu}), (\ref{eta}). The consequences of disregarding a term $\left<H^{(R)}\right>_{h^{(M)}}$ are discussed in section~\ref{Sec:CriticalMagneticField}.

The two equations (\ref{nu}) and (\ref{eta}) can be reduced to a single one for $\nu$. The only requirement is $g > 0$. The general form of Eqs.~(\ref{nu}), (\ref{eta}) is
\begin{equation}
\label{Eq:NuGen}
\nu = f_{1}(\nu - \eta) + f_{2}(\nu),
\end{equation}
\begin{equation}
\label{Eq:EtaGen}
\eta = f_{1}(\nu - \eta).
\end{equation}
Let $g > 0$, then $f_{2} > 0$. Furthermore,
\begin{equation}
\label{eta-nu}
\eta = \nu - f_{2}(\nu),
\end{equation}
which yields the equation for $\nu$:
\begin{equation}
\label{nudef}
\nu = f_{1}(f_{2}(\nu)) + f_{2}(\nu),
\end{equation}
where according to Eqs.~(\ref{Eq:NuGen}),~(\ref{Eq:EtaGen}):
\begin{equation}
\label{f1def}
f_{1}(\nu) = (M N\beta)^{-1} \frac{\partial}{\partial \nu} \ln \text{Tr} \exp[ - \beta \tilde{h}(\nu,0)],
\end{equation}
\begin{equation}
\label{f2def}
f_{2}(\nu) = (N\beta)^{-1} \frac{\partial}{\partial \nu} \ln \text{Tr} \exp[ - \beta h^{(1)}_{imp}].
\end{equation}

\section{Mean-field theory of $\tilde{h}$}
\label{Sec:MeanFieldDescription}
The form of the Hamiltonian $\tilde{h}$, given by Eq.~(\ref{hEl}) is analogous to
\begin{equation}
H_{\mathrm{BCS}}({\mathcal H}) = H_{0} + V_{\mathrm{BCS}} - \mu_{\mathrm{B}} {\mathcal H} \sigma_{z},
\end{equation}
describing a system of electrons with attractive BCS interaction in the presence of an external magnetic field $\mathcal H$ ($\mu_{\mathrm{B}}$ denotes the Bohr magneton).
The explicit form of the system's free energy  $F(h^{(M)}(\nu, \eta), \beta)$ can be therefore derived by exploiting the Bogolubov-Valatin transformation~\cite{Bogolyubov, Valatin} and the method developed in Ref.~\onlinecite{Rickayzen} for $H_{\mathrm{BCS}}({\mathcal H})$.

This has been done in Ref.~\onlinecite{DB11} for spin $S=1/2$ and $S=7/2$ magnetic impurities perturbing the BCS-superconductor. For further investigation, let us recall the final form of the free energy $F^{(S)}$ and briefly recapitulate the results of Ref.~\onlinecite{DB11}.

The free energy is given by the following equation
\begin{equation}
\label{F_1_2}
\begin{split}
F^{(S)} & = \min_{\{\Delta,\,\nu\}} \Bigl\{ \rho_{F} |\Lambda| \int_{-\delta}^{\delta} \Bigl[ \frac{1}{2} \Delta^{2} E^{-1} f_{3}\bigl(\beta, E, \xi, f_{2}^{(S)}\bigr) - \beta^{-1} \ln \bigl[2 \cosh(\beta E) + 2 \cosh\bigl(g \beta M f_{2}^{(S)}\bigr)\bigr]\Bigr]\mathrm{d}\xi \\ &+ M^{2}c^{-1} \bigl(\nu f_{2}^{(S)} - \frac{1}{2} \bigl(f_{2}^{(S)}\bigr)^{2}\bigr) + F_{imp}^{(S)} + E_{0}(\Delta = 0) + \rho_{F} \delta^{2}\Bigr\}, \qquad S = 1/2,\ 7/2,
\end{split}
\end{equation}
where $F_{imp}^{(S)}$ is the free energy of impurity subsystem, given by Eqs.~(\ref{Fimp_1_2}) and (\ref{Fimp_7_2}), $E = \sqrt{\Delta^{2} + \xi^{2}},$ $\rho_{F}$ denoting the density of states at Fermi level,
\[
\rho_{F} = \frac{m p_{F}}{2 \pi \hbar^{2}},
\]
whereas $E_{0}(\Delta = 0)$ denotes the ground state energy of free fermions. Two last terms in Eq.~(\ref{F_1_2}) are the contribution to the free energy density from one-fermion states, lying outside $\mathcal P$.

The system's state is characterized, according to equation~(\ref{F_1_2}), by the minimizing solution, $\{\Delta_{\mathrm{m}}, \nu_{\mathrm{m}}\}$, of the following set of equations for the gap $\Delta$ and a parameter $\nu$, describing the impurity subsystem
\begin{equation}
\label{Delta}
\Delta = \frac{1}{2} G_{0} \rho \int_{-\delta}^{\delta} \frac{\Delta}{E}f_{3}\bigl(\beta, E, \xi, f^{(S)}_{2}(\nu)\bigr)\mathrm{d}\xi \qquad S = 1/2,\ 7/2,
\end{equation}
\begin{equation}
\label{nuEq}
\nu = f_{1}\bigl(f^{(S)}_{2}(\nu)\bigr) + f^{(S)}_{2}(\nu) = \frac{cg}{M} \frac{\sinh\bigl(\beta g M f^{(S)}_{2}(\nu)\bigr)}{\cosh\bigl(\beta g M f^{(S)}_{2}(\nu)\bigr) + \cosh(\beta E_{\mathbf{k}})} + f^{(S)}_{2}(\nu),
\end{equation}
where
\begin{equation}
\label{f3}
f_{3}(\beta, E, \xi, f_{2}) = \frac{\sinh(\beta E)}{\cosh(\beta E) + \cosh(g \beta M f_{2}^{(S)}(\nu))}.
\end{equation}
The properties of a superconductor with magnetic impurities can be determined by solving this set of equations, which is supplemented by the following condition for the chemical potential $\mu$:
\begin{equation}
\label{ChemicalPotential}
\sum_{\mathbf{k}\sigma}\text{Tr}\bigl(n_{\mathbf{k}\sigma} \rho_{0} \bigr) = n,
\end{equation}
where $n_{\mathbf{k}\sigma} = a_{\mathbf{k}\sigma}^{\dagger}a_{\mathbf{k}\sigma}^{}$ is the fermion number operator and $n$ denotes the average number of fermions in the system. It has been shown in Ref.~\onlinecite{DB11}, that this condition, takes the form:
\begin{equation}
\label{mu}
\sum_{\mathbf{k}} \Bigl[1 - \frac{\xi_{\mathbf{k}}}{E_{\mathbf{k}}} f_{3}\bigl(\beta, E_{\mathbf{k}}, \xi_{\mathbf{k}}, f^{(S)}_{2}\bigr) \Bigr] = n.
\end{equation}
Equation~(\ref{mu}) resembles the BCS equation for $\mu$ and the properties of $f_{3}$ are similar to those of $f_{\mathrm{BCS}} = \tanh(\beta E_{\mathbf{k}}/2)$, e.g. both functions are odd in $\xi_{\mathbf{k}}$. The solution of equation~(\ref{mu}) is therefore exactly the same as in BCS theory, viz., $\mu = \varepsilon_{\mathrm{F}}$. Thus, we assume that in the low-temperature scale the following relations hold:
\begin{equation}
\mu = \varepsilon_{\mathrm{F}}, \qquad \frac{\partial \mu}{\partial T} = 0, \qquad \rho = \rho_{\mathrm{F}}.
\end{equation}

Equations (\ref{Delta}) and (\ref{nuEq}) clearly possess the solution $\Delta = \nu = 0$ for all values of $\beta \geq 0$. At sufficiently large values of $\beta$ one finds also other solutions, viz., $\{ \Delta \neq 0, \nu = 0 \}$, $\{ \Delta = 0, \nu \neq 0 \}$, $\{ \Delta \neq 0, \nu \neq 0 \}$. Accordingly, we distinguish the following phases:
\begin{itemize}
\item[$-$]{paramagnetic phase $P$ with $\{\Delta_{\mathrm{m}} = 0, \nu_{\mathrm{m}} = 0 \}$,}
\item[$-$]{unperturbed superconducting state $SC$ with $\{\Delta_{\mathrm{m}} \neq 0, \nu_{\mathrm{m}} = 0 \}$,}
\item[$-$]{ferromagnetic phase $F$ without bound Cooper pairs and $\{\Delta_{\mathrm{m}} = 0, \nu_{\mathrm{m}} \neq 0 \}$, in which impurity spins tend to align opposite to those of conduction fermions (cf. Eqs.~(\ref{hEl}) and (\ref{hImp})).}
\item[$-$]{intermediate phase $D$ in which superconductivity coexists with ferromagnetism and $\{\Delta_{\mathrm{m}} \neq 0, \nu_{\mathrm{m}} \neq 0 \}$.}
\end{itemize}

We define the following temperatures corresponding to the respective phase transitions
\begin{itemize}
\item[$-$]{$T_{\mathrm{c}}$, 2nd order transition $SC$ $\rightarrow$ $P$.}
\item[$-$]{$T_{PF}$, Curie temperature of 2nd order transition $F$ $\rightarrow$ $P$.}
\item[$-$]{$T_{SCD}$, 1st order transition $D$ $\rightarrow$ $SC$.}
\item[$-$]{$T_{FD}$, 1st order transition $D$ $\rightarrow$ $F$.}
\item[$-$]{$T_{SCF}$, 1st order transition $SC$ $\rightarrow$ $F$.}
\end{itemize}

The set of Eqs.~(\ref{Delta}),~(\ref{nuEq}) were solved numerically for different values of $g$, $M$, $\delta$ and $G_{0}\rho_{\mathrm{F}}$ and solutions, minimizing free energy are exploited in next section to determine the critical magnetic field. The parameters were adjusted to fit the experimental data for LaCe, ThGd and SmRh$_{4}$B$_{4}$. These results are discussed in section~\ref{Sec:CriticalMagneticField}.

The free energy (Eq.~\ref{F_1_2}), as well as the equations for $\Delta$ (Eq.~(\ref{Delta})) and $\nu$ (Eq.~(\ref{nuEq})) strongly depend on the value of the impurity spin $S$. In the present work we study the effect on superconductivity of the following magnetic ions: Ce ($S=1/2$), Gd and Sm ($S=7/2$). For $S=1/2$ one obtains
\begin{equation}
\label{Fimp_1_2}
\begin{split}
F_{imp}^{(\frac{1}{2})} &= -\beta^{-1} \ln \text{Tr} \exp \bigl[-\beta h_{imp}^{(M)}\bigr] = -\beta^{-1} \sum_{\alpha = 1}^{M} \ln \text{Tr} \exp \bigl[-\beta h_{imp}^{(1)}\bigr] \\&= - M \beta^{-1} \ln \bigl[2 \cosh(\beta g \nu)\bigr] + \frac{1}{2} c g^{2},
\end{split}
\end{equation}
with
\begin{equation}
\label{f2_1_2}
f_{2}^{(\frac{1}{2})}(\nu) = \frac{c g}{M} \tanh(\beta g \nu).
\end{equation}
Accordingly, for spin $7/2$ impurities
\begin{equation}
\label{Fimp_7_2}
\begin{split}
F_{imp}^{(\frac{7}{2})} = & - M \beta^{-1} \ln 2 \Bigl[\exp[-24 g^{2} \beta N^{-1}] \cosh(7 \beta g \nu) + \exp[-12 g^{2} \beta N^{-1}] \cosh(5 \beta g \nu) \\ &+ \exp[-4 g^{2} \beta N^{-1}] \cosh(3 \beta g \nu) + \cosh( \beta g \nu) \Bigr] + \frac{1}{2}cg^{2},
\end{split}
\end{equation}
where
\begin{equation}
\label{f2_7_2}
\begin{split}
f_{2}^{(\frac{7}{2})}(\nu) & = \frac{c g}{M R} \Biggl[ 7 \exp[-24g^{2} \beta N^{-1}] \sinh(7 \beta g \nu) + 5 \exp[-12 g^{2} \beta N^{-1}] \sinh(5 \beta g \nu) \\ & + 3\exp[-4 g^{2} \beta N^{-1}] \sinh(3 \beta g \nu) + \sinh(\beta g \nu) \Biggr],
\end{split}
\end{equation}
and
\[
\begin{split}
R &= \exp[-24 g^{2} \beta N^{-1}] \cosh(7 \beta g \nu) + \exp[-12 g^{2} \beta N^{-1}] \cosh(5 \beta g \nu) \\ &+ \exp[-4 g^{2} \beta N^{-1}] \cosh(3 \beta g \nu)  + \cosh( \beta g \nu).
\end{split}
\]
The complexity of the free energy $F^{(S)}$ and functions $f_{1}$, $f_{2}$, $f_{3}$ increases with the impurity spin value. It follows, that the impurity spin is the key factor affecting the thermodynamics of superconducting magnetic alloys. This conclusion is complementary with the fundamental experimental observation made by Matthias et al.~\cite{Matthias1}.

\section{The critical magnetic field}
\label{Sec:CriticalMagneticField}
The critical magnetic field $H_{\mathrm{c\Phi}}$ forcing a system to undergo the phase transition from the $\Phi$ phase to paramagnetic (normal) phase $P$ is given by the equation
\begin{equation}
\label{SquaredCriticalField}
H_{\mathrm{c}\Phi}^{2} = (F_{P} - F_{\Phi})/2 \mu_{0},
\end{equation}
where $\mu_{0}$ denotes the vacuum permeability, $F_{P}$ and $F_{\Phi}$ denote the free energy of the $P$ and $\Phi$ phase respectively.

The free energy of the normal state can be obtained from equation (\ref{F_1_2}) with $\Delta = 0$ and $\nu = 0$, which yields
\begin{itemize}
\item[$-$]{for spin $1/2$ impurities
\begin{equation}
\begin{split}
F_{P}^{(\frac{1}{2})} = & -2\,\rho_{F}^{}|\Lambda| \beta^{-1} \int_{-\delta}^{\delta} \ln 2\cosh(\frac{1}{2}\beta \xi) \mathrm{d}\xi - M \beta^{-1} \ln 2 + \frac{1}{2}c g^2 \\& + E_{0}(\Delta = 0) + \rho_{F}^{} \delta^{2},
\end{split}
\end{equation}
}
\item[$-$]{for spin $7/2$ impurities
\begin{equation}
\begin{split}
F_{P}^{(\frac{7}{2})} =&  -2\,\rho_{F}^{}|\Lambda| \int_{-\delta}^{\delta} \ln 2\cosh(\frac{1}{2}\beta \xi) \mathrm{d}\xi - M \beta^{-1} \ln 2 \Bigl[\exp[-24 \beta c g^2 M^{-1}] \\& - \exp[-12 \beta c g^2 M^{-1}] - \exp[-4 \beta c g^2 M^{-1}] + 1 \Bigr] + \frac{1}{2}c g^2 \\ & + E_{0}(\Delta = 0) + \rho_{F}^{} \delta^{2}.
\end{split}
\end{equation}
}
\end{itemize}
The critical magnetic field required to suppress superconductiviy (phase transition $SC \rightarrow P$) is given by the equation (\ref{SquaredCriticalField}) with $F_{SC}$ replacing $F_{\Phi}$. It will be denoted as usual by $H_{\mathrm{c}}$.
The expression for $F_{SC}$ results from equation (\ref{F_1_2}) with $\{\Delta \neq 0, \nu = 0 \}$:
\begin{equation}
\begin{split}
F_{SC}^{(\frac{1}{2})} = &\rho_{F}^{}|\Lambda| \int_{-\delta}^{\delta} \Bigl[\frac{1}{2} \Delta^{2} E^{-1} f_{3}^{(BCS)}(\beta, E, \xi) - 2 \beta^{-1} \ln 2 \cosh(\beta E)\Bigr]\mathrm{d}\xi \\ & + \frac{1}{2}c g^2 + E_{0}(\Delta = 0) + \rho_{F}^{} \delta^{2}, \qquad \mathrm{for}\ S = 1/2,
\end{split}
\end{equation}
and
\begin{equation}
\begin{split}
F_{SC}^{(\frac{7}{2})} = & \rho_{F}^{} |\Lambda| \int_{-\delta}^{\delta} \Bigl[\frac{1}{2} \Delta^{2} E^{-1} f_{3}^{(BCS)}(\beta, E, \xi) - 2 \beta^{-1} \ln 2 \cosh(\beta E)\Bigr]\mathrm{d}\xi \\ & - M \beta^{-1} \ln 2 \Bigl[\exp[-24 \beta c g^2 M^{-1}] - \exp[-12 \beta c g^2 M^{-1}] - \exp[-4 \beta c g^2 M^{-1}] + 1 \Bigr] \\& + \frac{1}{2}c g^2 + E_{0}(\Delta = 0) + \rho_{F}^{} \delta^{2}, \qquad \mathrm{for }\ S = 7/2.
\end{split}
\end{equation}

The credibility of the above theoretical expressions will be now verified on the expertimental data of LaCe, ThGd and SmRh$_{4}$B$_{4}$. The parameters $g,\ G_{0}\rho_{F},$ $\delta,\ M$ were adjusted to provide the best possible fit of the theoretical critical magnetic curves to the experimental data. In order to perform a fit, the set of equations~(\ref{Delta}), (\ref{nuEq}) was solved numerically for fixed values of the parameters $g,\ \rho_{F}|\Lambda|,\ G_{0}\rho_{F},\ \delta,\ M$. The minimizing solution of this set of equations was subsequently substituted to the expressions for $H_{\mathrm{c}}$ or $H_{\mathrm{c}D}$ depending on which of the system's states ($SC$ or $D$) possess the smaller values of the free energy, i.e. which of them is thermodynamically stable.

\begin{figure}
\begin{tabular}{@{}c@{ }c@{ }c@{ }c@{}@{ }@{ }c@{ }c@{ }c@{ }c@{ }}
\multicolumn{1}{l}{\footnotesize \bf a} & \multicolumn{1}{l}{\footnotesize{\bf b}}\\[-1.0cm]
    \includegraphics[width=0.5\textwidth]{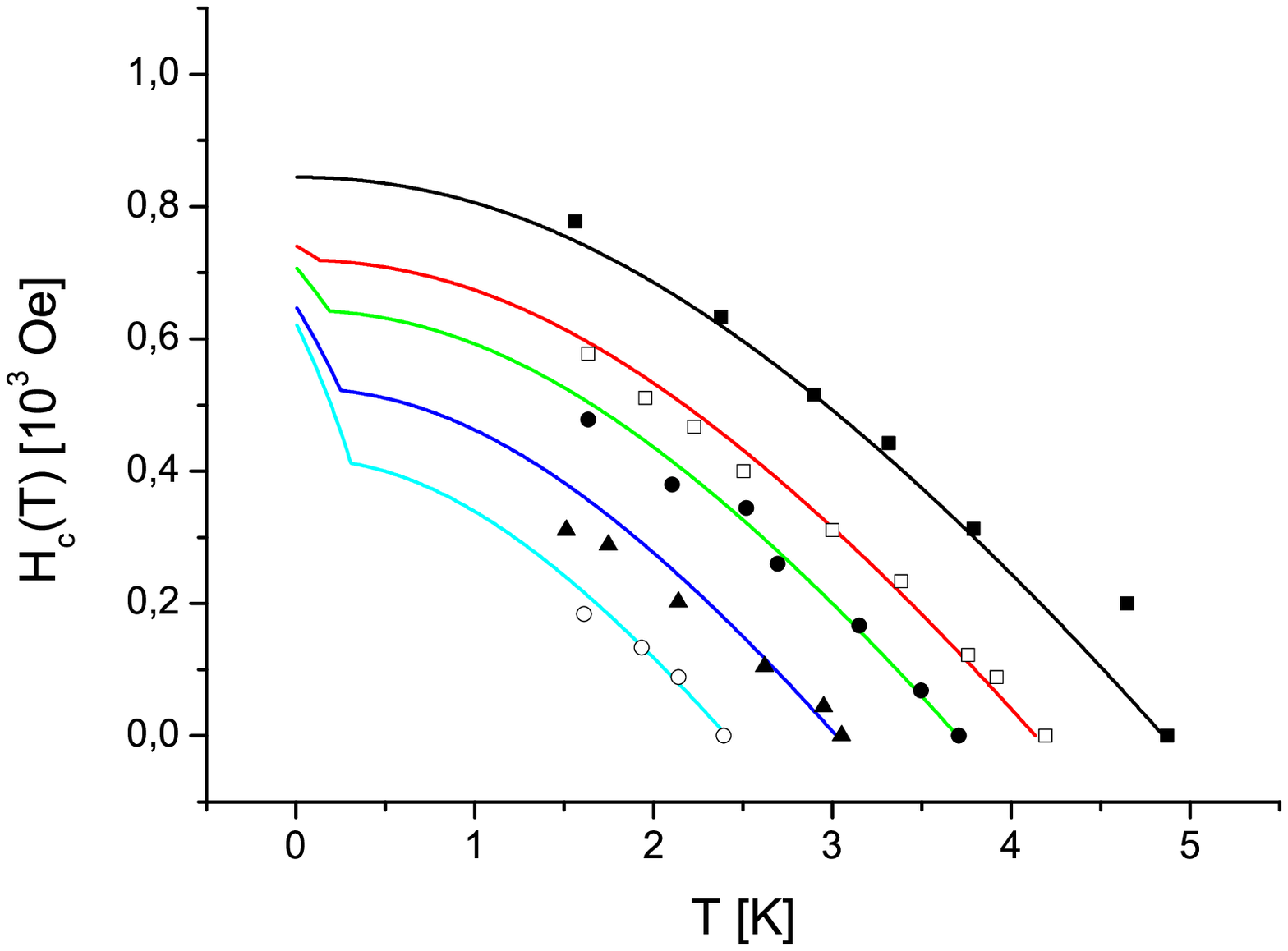}&
    \includegraphics[width=0.5\textwidth]{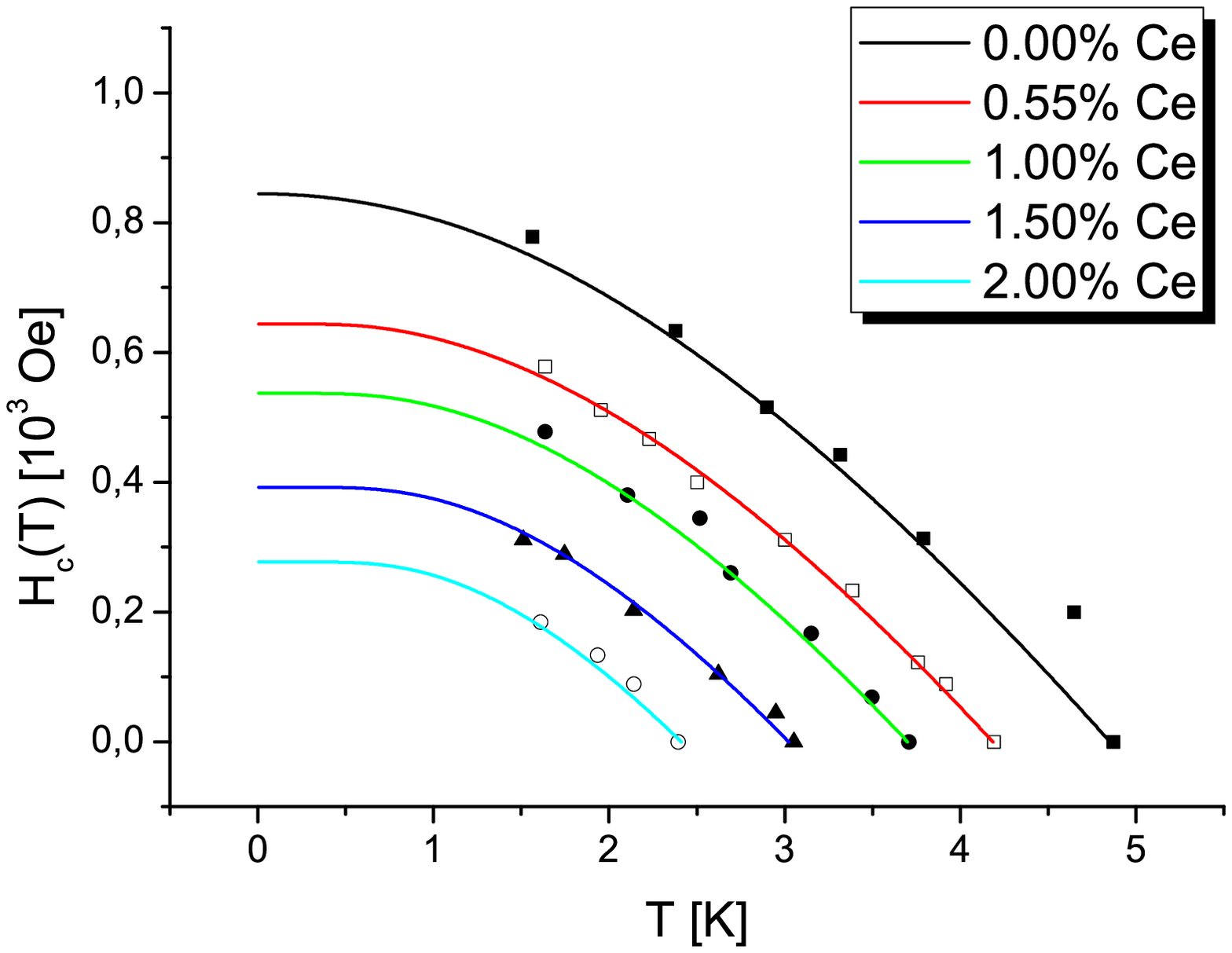}
\end{tabular}
\caption{\label{hcLaCe} The critical magnetic field of LaCe under varying Ce concetration for real temperature $T$ and parameter values given in Table~\ref{table3} (a) and for the effective temperature $\tilde{T}$ and parameter values given in Table~\ref{table4} (b). The points are experimental data from Ref.~\onlinecite{SugawaraEguchi}.}
\end{figure}

\begin{figure}
\begin{center}
\begin{tabular}{@{}c@{ }c@{ }c@{ }c@{}@{ }@{ }c@{ }c@{ }c@{ }c@{ }}
\multicolumn{1}{l}{\footnotesize \bf a} & \multicolumn{1}{l}{\footnotesize{\bf b}}\\[-1cm]
    \includegraphics[width=8cm]{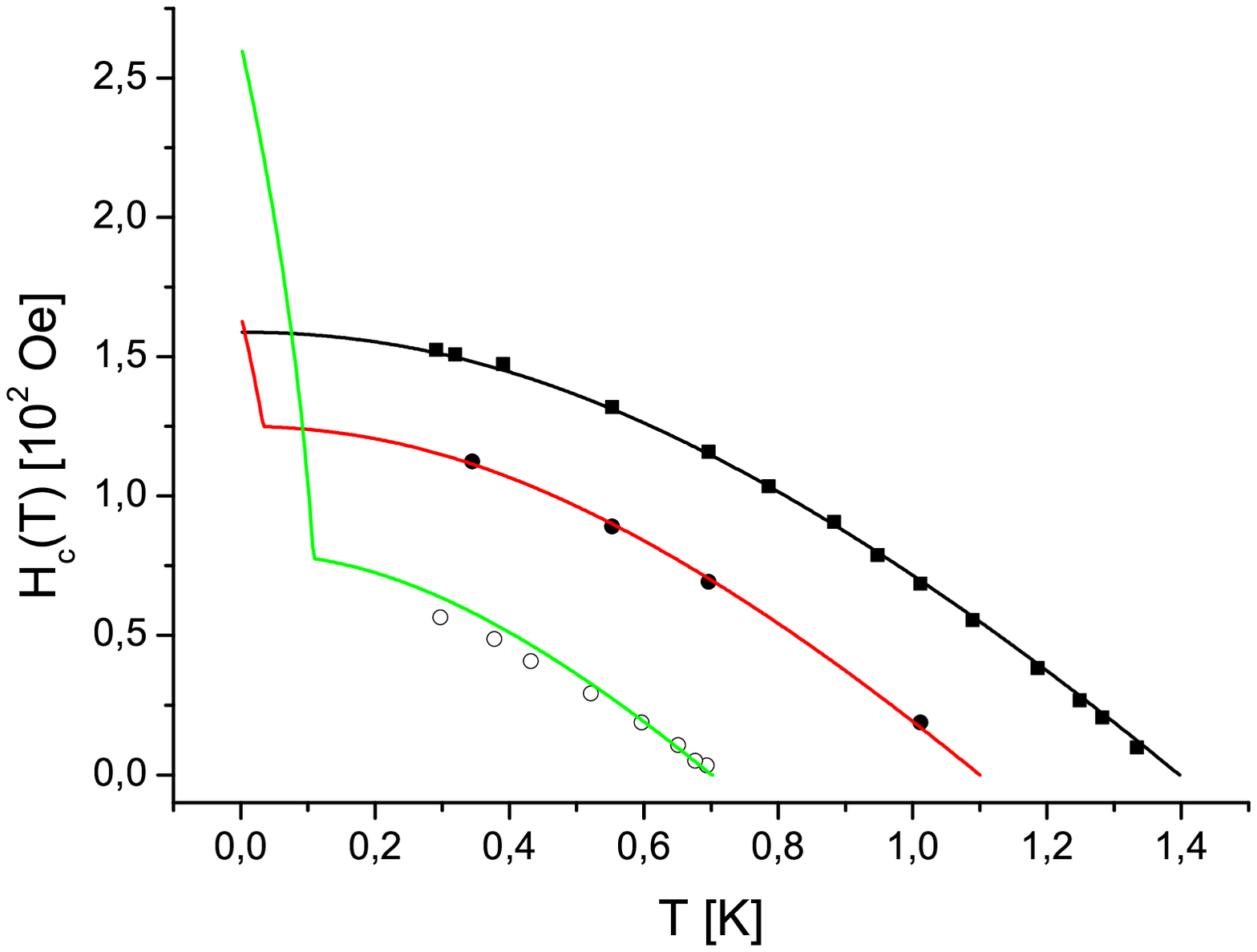}&
    \includegraphics[width=8cm]{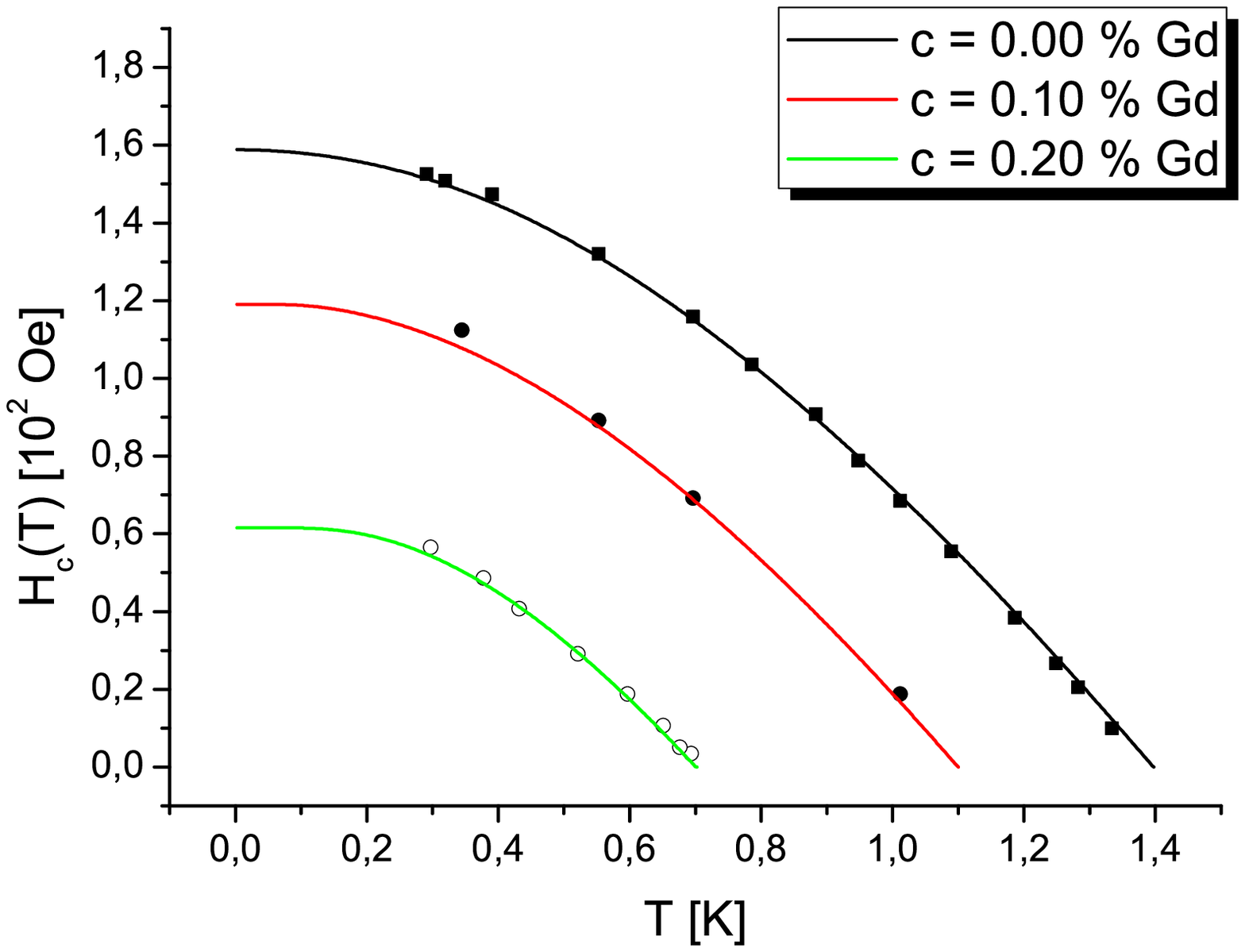}\\[-1em]
\end{tabular}
\end{center}
\caption{\label{hcThGd} The critical magnetic field of ThGd under varying Gd concetration for: real temperature $T$ and parameter values given in Table~\ref{table3} (a) the effective temperature $\tilde{T}$ and parameter values given in Table~\ref{table4} (b). The points are experimental data from Ref.~\onlinecite{DeckerPetersonFinnemore}.}
\end{figure}

\begin{figure}
\begin{center}
	\includegraphics[width=8cm]{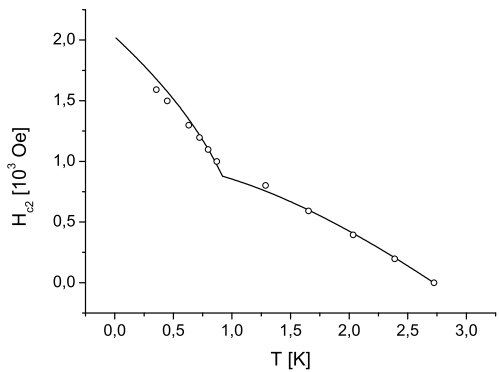}
\end{center}
\caption{\label{hcSmRh4B4} The upper critical magnetic field $H_{\mathrm{c}2}$ of SmRh$_{4}$B$_{4}$ for the parameter values given in Table~\ref{table3}. The points are experimental data from Ref.~\onlinecite{HWMFM78}. $H_{\mathrm{c}2}$ is related to the thermodynamic critical magnetic field $H_{\mathrm{c}}$ by the expression, $H_{\mathrm{c}2} = \kappa \sqrt{2} H_{\mathrm{c}}$. The Ginzburg-Landau parameter $\kappa$ has been treated as the additional adjustable parameter used to fit experimental data and assumed to be independent in $T$. The value of $\kappa$ providing best fit is $\kappa = 1.51$.}
\end{figure}

The critical magnetic field of LaCe and ThGd are depicted in Figs.~\ref{hcLaCe}a, ~\ref{hcThGd}a and \ref{hcSmRh4B4}. The agreement with experiment is satisfactory. The discrepancies between theoretical and experimental data increases with decreasing temperature and increasing impurity concentration in case of LaCe and ThGd. However, these discrepancies can be reduced by introducing the concept of the effective inverse temperature $\tilde{\beta} = 1 / k_{\mathrm{B}}\tilde{T}$, which is related to the system's real temperature by the following formulae
\begin{equation}
\label{EffectiveTemperature}
\tilde{\beta}(\beta, \gamma) = \gamma^{-1} \tanh(\beta \gamma).
\end{equation}
The effective temperature results by averaging (over the impurity positions) the single particle equilibrium density matrix of a quantum particle in a field of randomly positioned wells, representing the screened Coulomb potential at each impurity site \cite{Mackowiak}. Furthermore, it has been shown, that the $\gamma$ is of the form
\[
\gamma = \frac{\hbar}{2}\sqrt{M u_{2} m^{-1}},
\]
with $u_{2}$ denoting the 2nd derivative at well's minimum. In the present work $\gamma$ will be treated as the adjustable parameter.

Theoretical curves of the critical magnetic field of LaCe and ThGd superconducting alloys with $\tilde{\beta}$ replacing $\beta$ in Eq.~(\ref{SquaredCriticalField}) are given in Figs.~\ref{hcLaCe}b,~\ref{hcThGd}b. It is clear, that the application of the effective temperature improves the agreement of the given model with experimental data for LaCe and ThGd, proving that the Coulomb interactions apart from Coulomb interactions are also important factor of the superconducting alloys theory. The effective temperature has been also proven to be the crucial in description of the doping dependence of superconducting transition temperature $T_{\mathrm{c}}$ in high-$T_{\mathrm{c}}$ superconductors \cite{MB10}.

In case of SmRh$_{4}$B$_{4}$, the concentration of magnetic impurities (Sm$^{3+}$ ions), $c \approx 11$\% is much larger than in case of LaCe and ThGd. Accordingly, the agreement with experimental data for SmRh$_{4}$B$_{4}$ is only satisfactorly and can not be improved by the application of the effective temperature. This suggests, that the term $\left<H^{(R)}\right>_{h^{(M)}}$ can not be disregarded for sufficiently large $c$ and shall be included in the free energy $F(h^{(M)}, \beta)$ computation or the dependence of the Ginzburg-Landau parameter $\kappa$ on $T$ shall be taken into account. In the case of large $c$, the magnitude of exchange interaction between conduction fermions and magnetic ions may presumably exceeds the magnitude of the Coulomb attraction between magnetic ions and conduction fermions.

\begin{figure}
\begin{center}
\begin{tabular}{@{}c@{ }c@{ }c@{ }c@{}@{ }@{ }c@{ }c@{ }c@{ }c@{ }}
\multicolumn{1}{l}{\footnotesize \bf a} & \multicolumn{1}{l}{\footnotesize{\bf b}}\\[-1cm]
    \includegraphics[width=8cm]{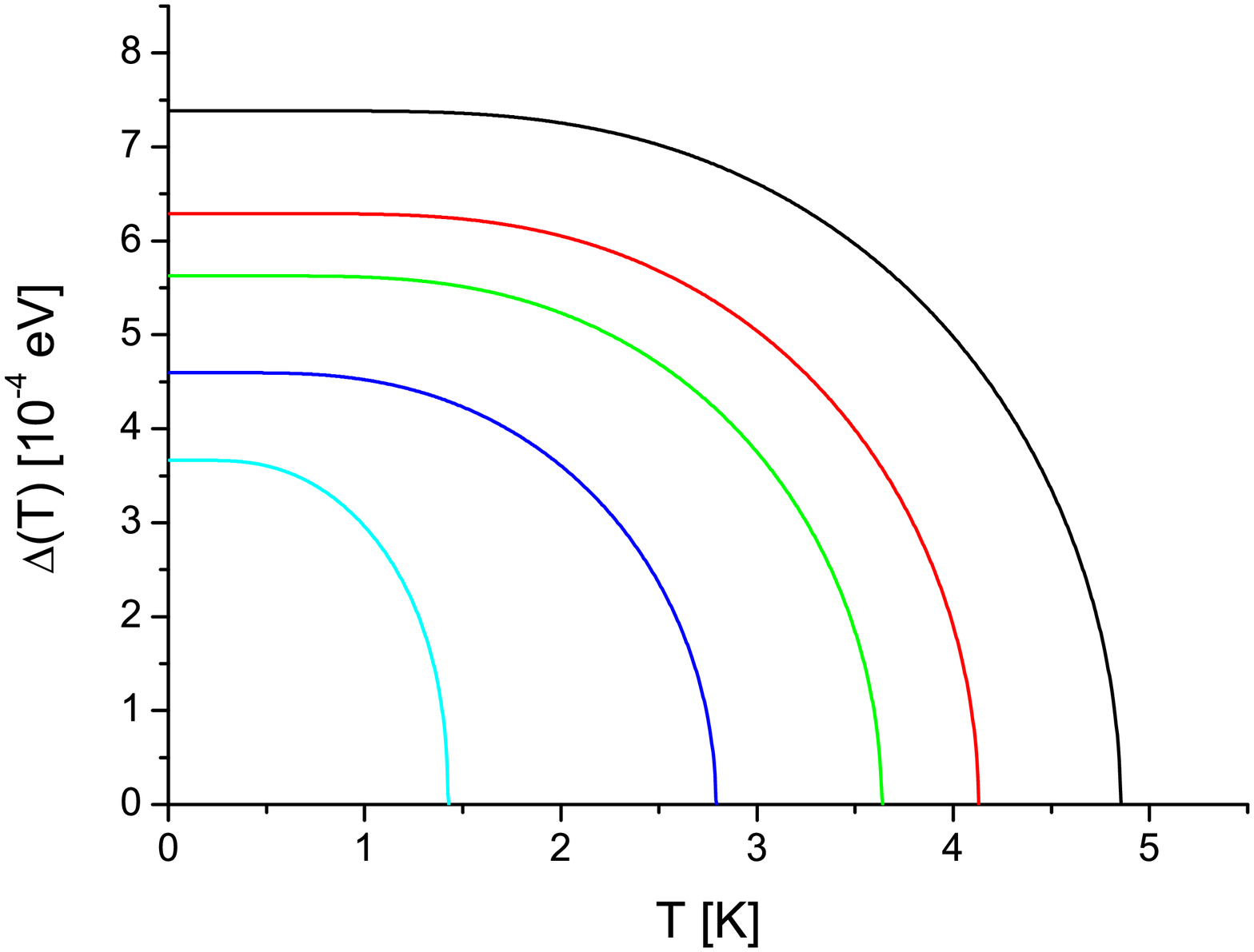}&
    \includegraphics[width=8cm]{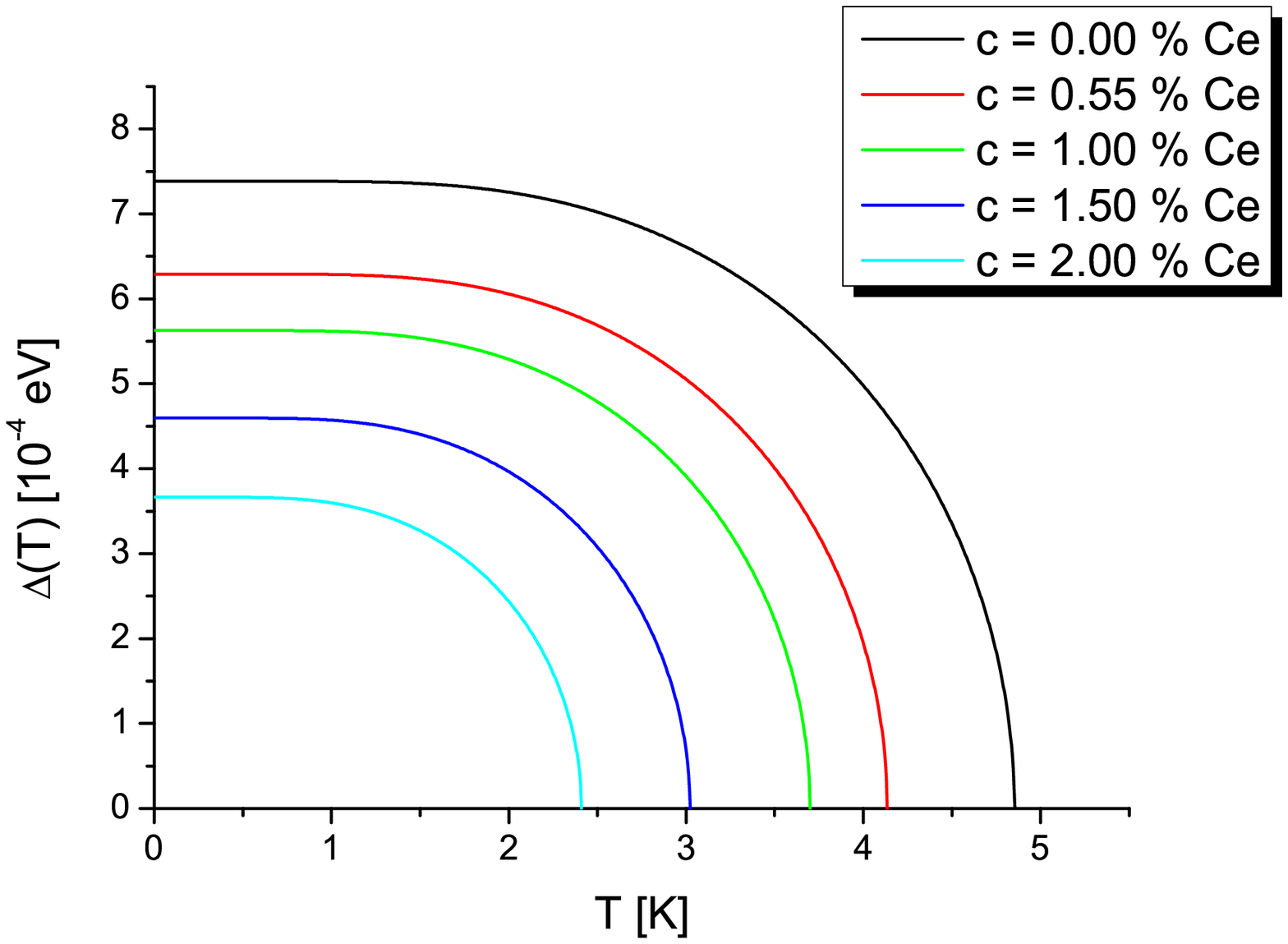}\\[-1em]
\end{tabular}
\end{center}
\caption{\label{f:DeltaLaCe} The temperature dependence of the gap parameter $\Delta(T)$ for LaCe and the solution of Eqs.~(\ref{Delta}),~(\ref{nuEq}) with $S=1/2$, in which $\{\Delta \neq 0, \nu = 0\}$ (a) and $\{\Delta \neq 0, \nu \neq 0\}$ (b). The parameter values are given in Table~\ref{table3}.}
\end{figure}

\begin{figure}
\begin{center}
\begin{tabular}{@{}c@{ }c@{ }c@{ }c@{}@{ }@{ }c@{ }c@{ }c@{ }c@{ }}
\multicolumn{1}{l}{\footnotesize \bf a} & \multicolumn{1}{l}{\footnotesize{\bf b}}\\[-1cm]
    \includegraphics[width=8cm]{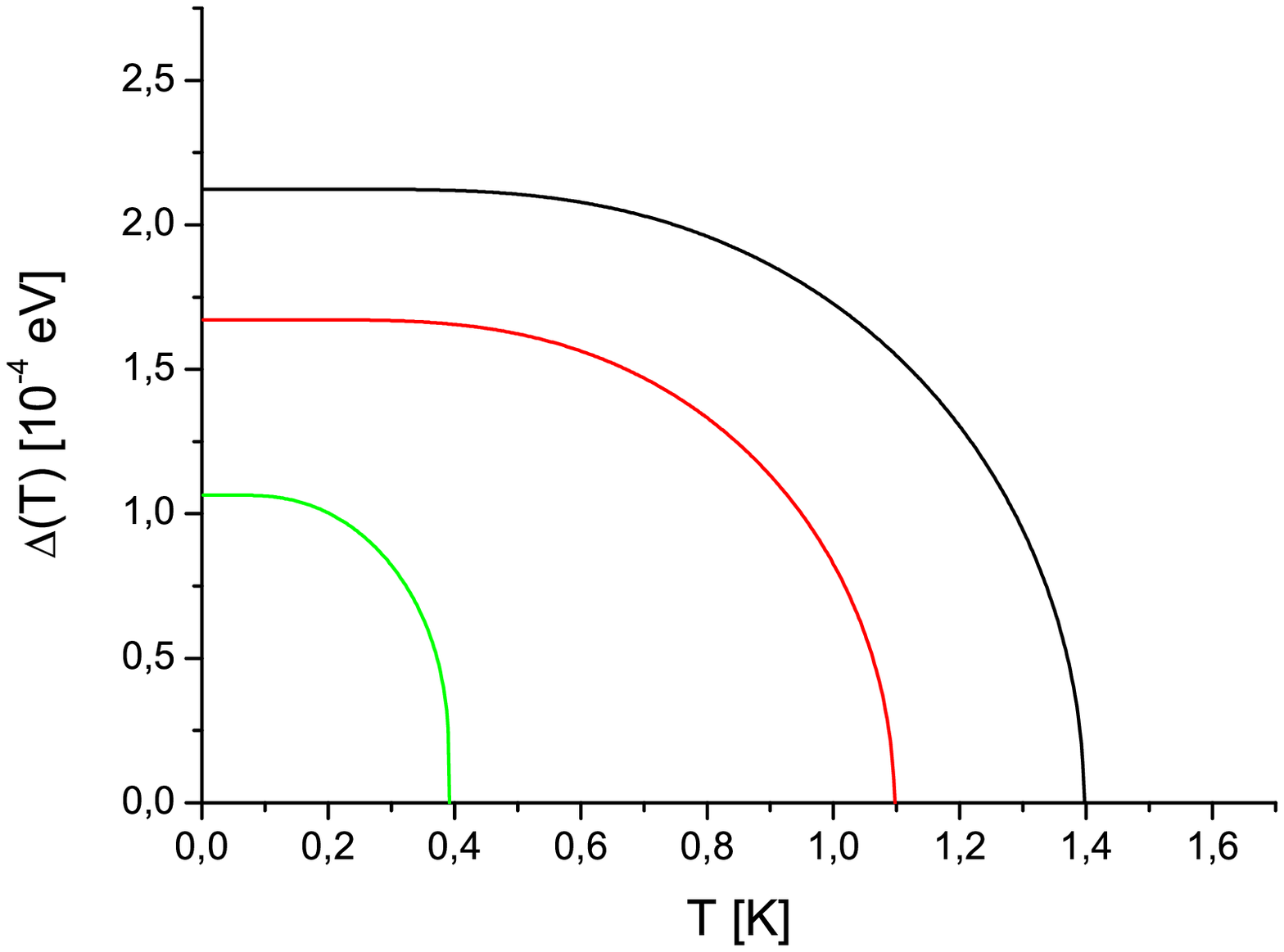}&
    \includegraphics[width=8cm]{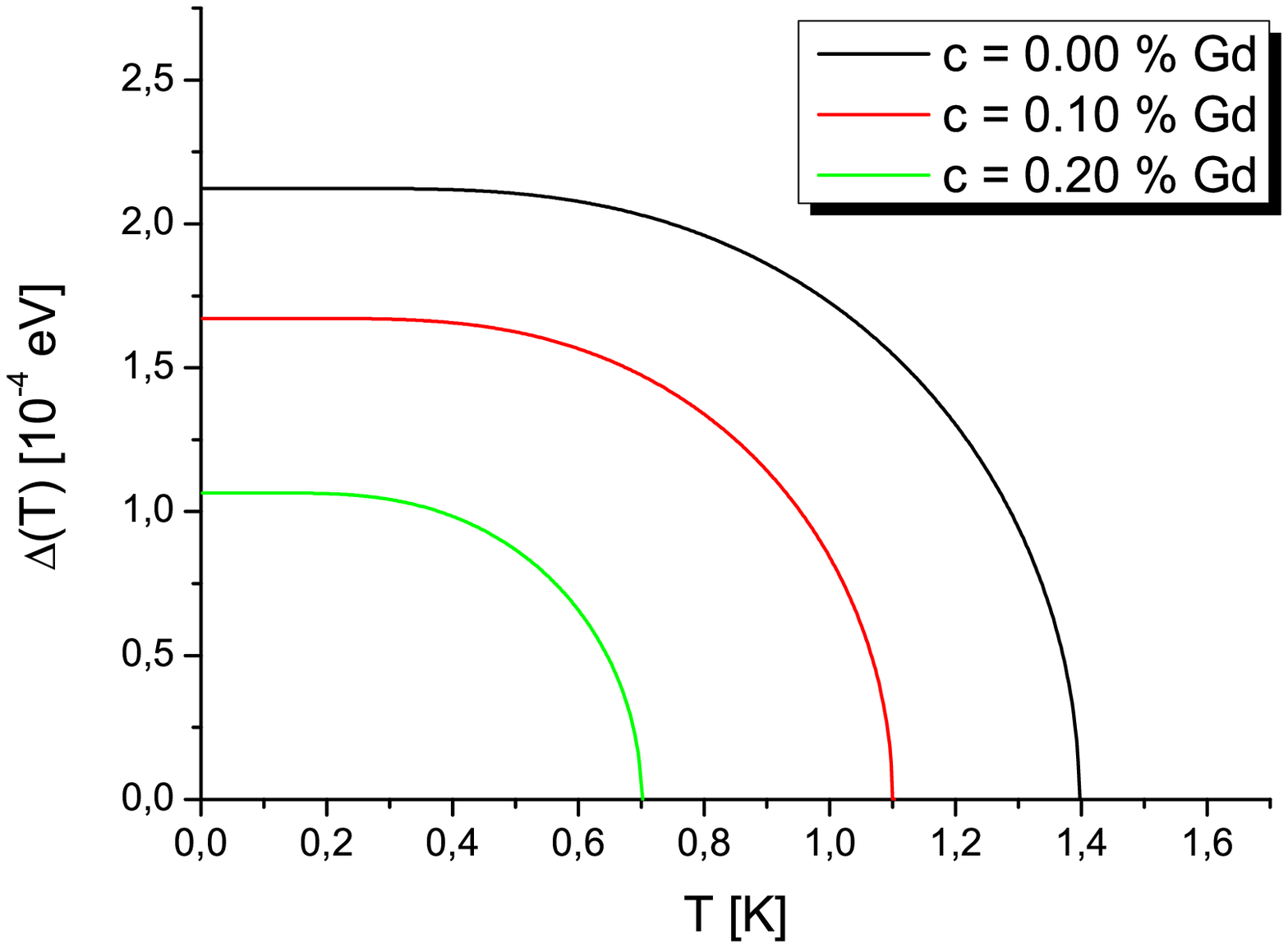}\\[-1em]
\end{tabular}
\end{center}
\caption{\label{f:DeltaThGd} The temperature dependence of the gap parameter $\Delta(T)$ for ThGd and the solution of Eqs.~(\ref{Delta}),~(\ref{nuEq}) with $S=7/2$, in which $\{\Delta \neq 0, \nu \neq 0\}$ (a) and $\{\Delta \neq 0, \nu = 0\}$ (b). The parameter values are given in Table~\ref{table3}.}
\end{figure}

\begin{figure}
\begin{center}
\begin{tabular}{@{}c@{ }c@{ }c@{ }c@{}@{ }@{ }c@{ }c@{ }c@{ }c@{ }}
\multicolumn{1}{l}{\footnotesize \bf a} & \multicolumn{1}{l}{\footnotesize{\bf b}}\\[-1cm]
    \includegraphics[width=8cm]{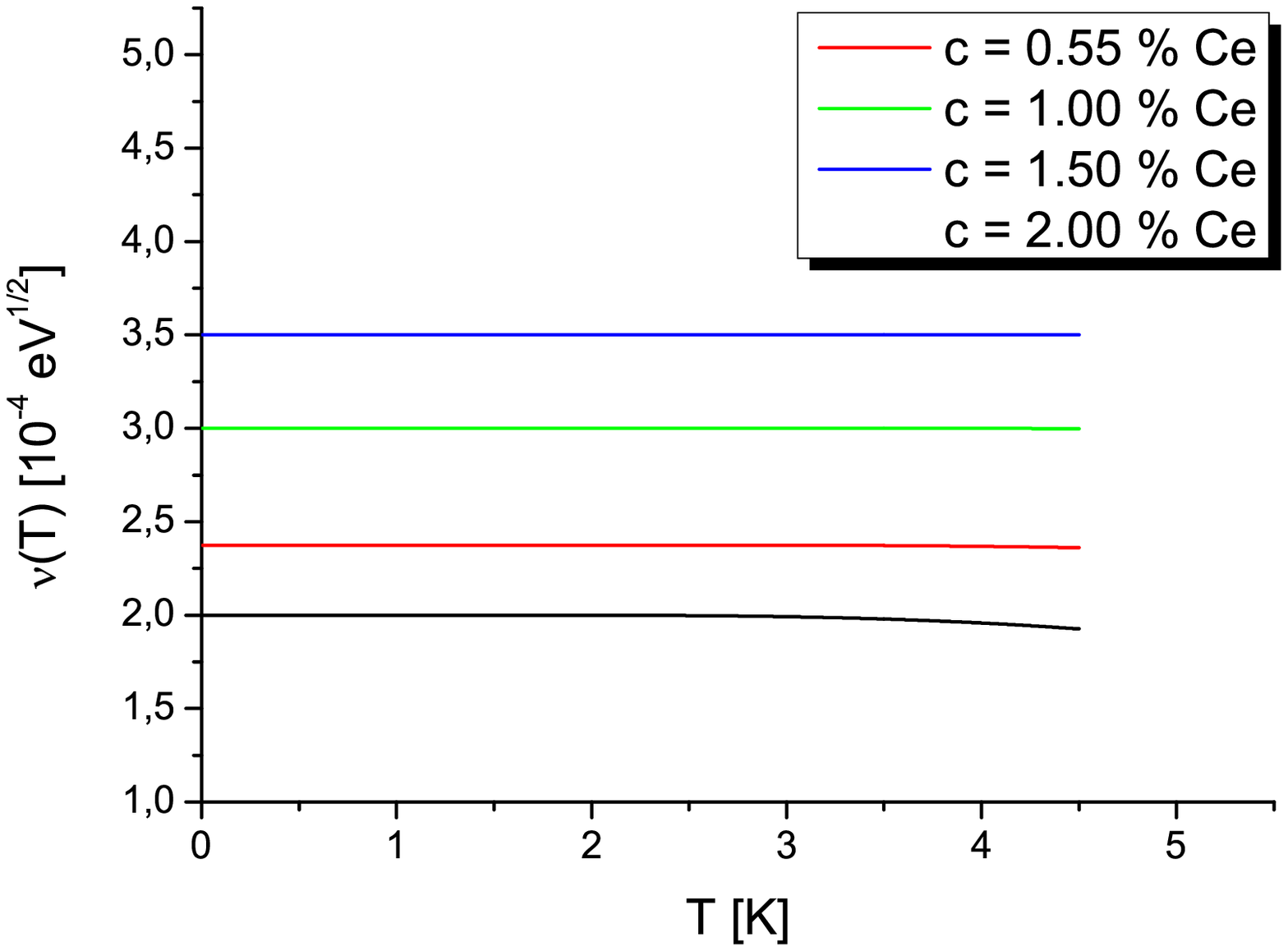}&
    \includegraphics[width=8cm]{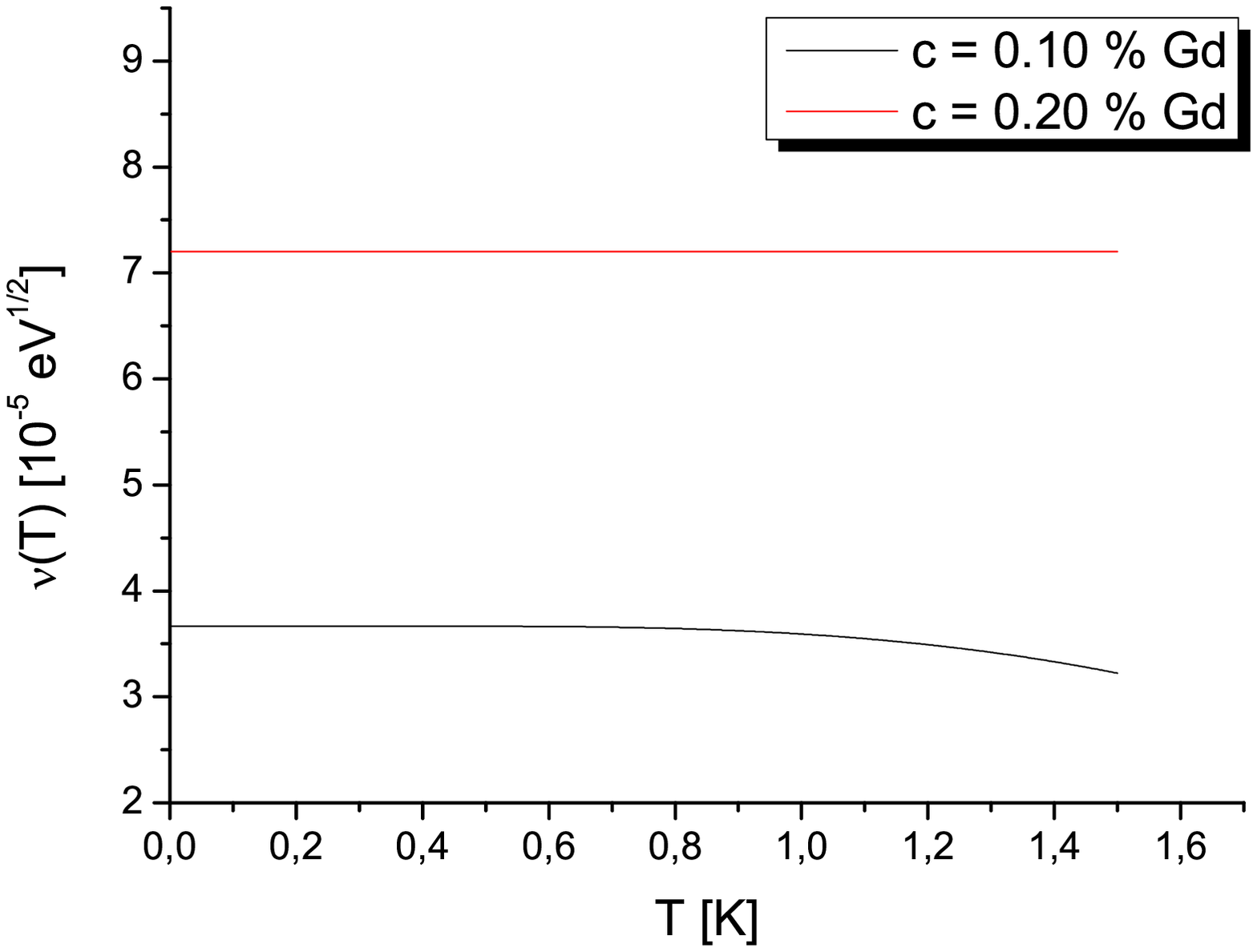}\\[-1em]
\end{tabular}
\end{center}
\caption{\label{f:Nu} The temperature dependence of the parameter $\nu(T)$ for the solution of Eqs.~(\ref{Delta}),~(\ref{nuEq}), in which $\{\Delta \neq 0, \nu \neq 0\}$ for LaCe (a) and ThGd (b). The parameter values are given in Table~\ref{table3}.}
\end{figure}

\begin{figure}
\begin{center}
\begin{tabular}{@{}c@{ }c@{ }c@{ }c@{}@{ }@{ }c@{ }c@{ }c@{ }c@{ }}
\multicolumn{1}{l}{\footnotesize \bf a} & \multicolumn{1}{l}{\footnotesize{\bf b}}\\[-1cm]
    \includegraphics[width=8cm]{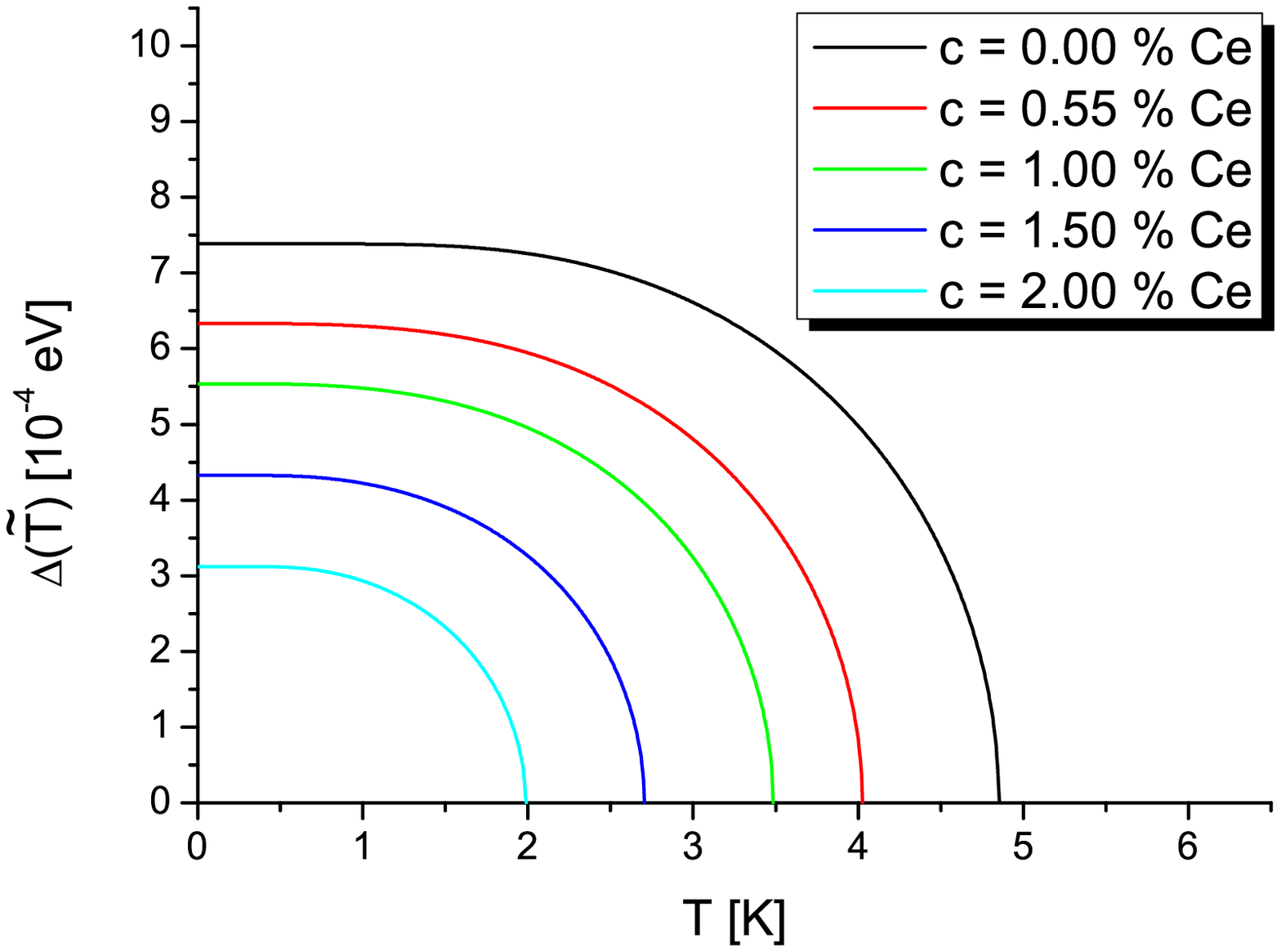}&
    \includegraphics[width=8cm]{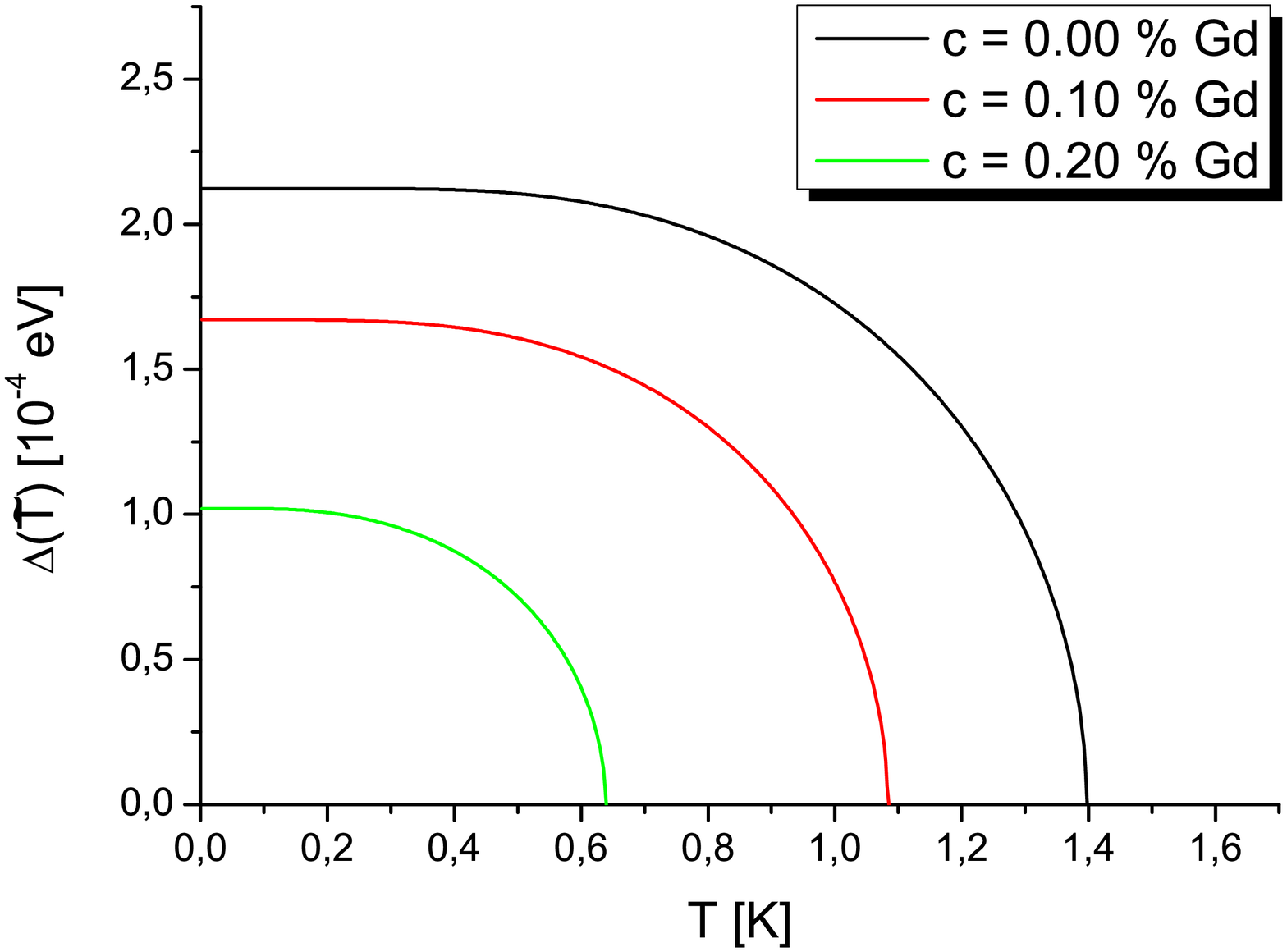}\\[-1em]
\end{tabular}
\end{center}
\caption{\label{f:DeltaTeff} The gap parameter $\Delta(\tilde{T})$ at the effective temperature $\tilde{T}$ and under varying impurity concentration for LaCe (a) and ThGd (b). The parameter values are given in Table~\ref{table4}.}
\end{figure}

The solutions of the set of equations for $\Delta$ and $\nu$ under varying temperature and impurity concentration of Ce and Gd are depicted in Figs.~\ref{f:DeltaLaCe}--\ref{f:DeltaTeff}.

The graphs, depicted in Figs.~\ref{hcLaCe}a and~\ref{hcThGd}a show an increase of the critical magnetic field at very low temperature scale and for sufficiently large values of impurity concentrations, e.g. for $c = 2.0 \%$ Ce at Fig.~\ref{hcLaCe}a and for $c = 0.20 \%$ Gd at Fig.~\ref{hcThGd}).

This enhancement of superconductivity also increases with impurity spin and impurity concentration, since at extremely low temperatures, the following formula
\begin{equation}
\label{HCD>}
H_{\mathrm{cD}}(c = 0.20 \% \mathrm{ Gd}) > H_{\mathrm{cD}}(c = 0.10\%\mathrm{ Gd})
\end{equation}
is valid.

The above observation results from the following fact. If the solution $\{\Delta \neq 0, \nu \neq 0\}$ minimizes the free energy, then the values of $H_{\mathrm{c}D}$ are larger than $H_{\mathrm{c}}$. Accordingly, one obtains an increase of the critical magnetic field below $T_{SCD}$, since at this temperature, the system undergoes a phase transition $SC \rightarrow D$ and $H_{\mathrm{c}D}$ becomes equal to $H_{\mathrm{c}}$. This conclusion may be recognized as incompatible with physical intuition, suggesting that the external magnetic field should gain the perturbative effect of magnetic impurities. As a result $H_{\mathrm{c}D}$ should possess smaller values than $H_{\mathrm{c}}$.

An increase of the upper critical magnetic field $H_{\mathrm{c}2} = \kappa \sqrt{2} H_{\mathrm{c}}$ with $\kappa = \lambda/\xi$, denoting the Ginzburg-Landau parameter, has been observed in the following materials: SmRh$_{4}$B$_{4}$, GdMo$_{6}$S$_{8}$, TbMo$_{6}$S$_{8}$, Sn$_{1-x}$Eu$_{x}$Mo$_{6}$S$_{8}$, Pb$_{1-x}$Eu$_{x}$Mo$_{6}$S$_{8}$, La$_{1.2-x}$Eu$_{x}$Mo$_{6}$S$_{8}$ \cite{OF82, BM82}. These experiments confirm the validity of inequality (\ref{HCD>}).

Fischer et al.~\cite{OF75} has pointed out that the superconductivity enhancement, represented by an increase of the critical magnetic field is a result of the Jaccarino-Peter effect\cite{VJ62}. This phenomenon may occur in the II-type superconductor, in which the magnetic moments of the impurities are antiferromagnetically coupled to that of the conduction fermions. This interaction generates an exchange field ${\mathcal H}_{\mathrm{J}}$, which acts on the spins of conduction electrons equivalently to an applied magnetic field, viz., breaks the Copper pairs. However, the negative sign of the coupling between the magnetic moments and the conduction fermions spins, determines the direction of ${\mathcal H}_{\mathrm{J}}$ to be opposite to that of $\mathcal H$. Thus, an applied magnetic field will be compensated by an exchange field, since the net magnetic field $\mathcal H_{\mathrm{T}}$ is given by ${\mathcal H} - |{\mathcal H}_{\mathrm{J}}|$. A given compound displays superconducting properties as long as the following relation holds
\begin{equation}
\label{Eq:HPlimit}
-{\mathcal H}_{\mathrm{p}} \leq {\mathcal H}_{\mathrm{T}} \leq {\mathcal H}_{\mathrm{p}},
\end{equation}
where
\begin{equation}
\label{Eq:Hp}
{\mathcal H}_{\mathrm{p}} = \sqrt{\frac{\rho_{F}^{}}{\chi_{\mathrm{P}} - \chi_{\mathrm{SC}}}}.
\end{equation}
$\chi_{\mathrm{P}}$ and $\chi_{\mathrm{SC}}$ denotes the magnetic susceptibility of the normal and superconducting state. ${\mathcal H}_{\mathrm{p}}$, defined by Eq.~(\ref{Eq:Hp}) is the Chandrasekhar-Clogston limiting paramagnetic field\cite{BC62,AC62}.

The Jaccarino-Peter effect has been observed experimentally in Eu$_{0.75}$Sn$_{0.25}$Mo$_{6}$S$_{7.2}$Se$_{0.8}$\cite{HM84}, during the investigation of the upper critical magnetic field. In particular, at low temperature scale, three subsequent phase transitions has been observed with increasing value of an external magnetic field, i.e. $SC \rightarrow P \rightarrow SC \rightarrow P$.

In conclusion, it is worth to point out, that the magnetic impurities were proven to limit the superconductivity, but on the other hand, under some specific conditions, they help the superconducting system to overcome the destructive effect of an external magnetic field. Furthermore, the interplay between superconductivity and magnetism is believed to be a possible mechanism of high-$T_{\mathrm{c}}$ superconductivity~\cite{BK04}, since the undoped state of cuprate superconductors is a strongly insulating antiferromagnet. The existence of such a parent correlated insulator is viewed to be an essential feature of high temperature superconductivity.

In the above discussion we dealt only with the critical magnetic field. To fully judge, if the Jaccarino-Peter compensation may occur in the superconducting alloys, one should study the effect of external magnetic field ${\mathcal H}$ on such system. This will be done in the next two section, in which we first study the effect of ${\mathcal H}$ on the superconducting alloys and then apply the resulting formulas to investigate the dependence of the superconducting critical temperature $T_{\mathrm{c}}$ on ${\mathcal H}$.

\begin{table}
\tabcolsep16pt
\begin{center}
\begin{tabular}{*{7}{|c}|}
\hline
Alloy & $c\ [\%]$ & $M$ & $g\ [\sqrt{\mathrm{eV}}]$ & $\delta\ [\mathrm{eV}]$ & $G_{0} \rho_{F}^{}$ \\
\hline
LaCe & $0.00$ & $0$ & $0.000$ & $0.01$ & $0.3030$ \\
LaCe & $0.55$ & $2$ & $0.080$ & $0.01$ & $0.2890$ \\
LaCe & $1.00$ & $4$ & $0.095$ & $0.01$ & $0.2800$ \\
LaCe & $1.50$ & $5$ & $0.100$ & $0.01$ & $0.2650$ \\
LaCe & $2.00$ & $6$ & $0.105$ & $0.01$ & $0.2500$ \\
\hline\hline
ThGd & $0.00$ & $0$ & $0.00$ & $0.01$ & $0.2200$ \\
ThGd & $0.10$ & $3$ & $0.11$ & $0.01$ & $0.2090$ \\
ThGd & $0.20$ & $5$ & $0.18$ & $0.01$ & $0.1910$ \\
\hline\hline
SmRh$_{4}$B$_{4}$ & $11.11$ & $7$ & $0.0625$ & $0.0106$ & $0.2540$\\
\hline
\end{tabular}
\caption{\label{table3} The parameter values exploited during the numerical analysis of Eqs.~(\ref{Delta}), (\ref{nuEq}) and (\ref{SquaredCriticalField}).}
\end{center}
\end{table}

\begin{table}
\tabcolsep10pt
\begin{center}
\begin{tabular}{*{7}{|c}|}
\hline
Alloy & $c\ [\%]$ & $M$ & $g\ [\sqrt{\mathrm{eV}}]$ & $\delta\ [\mathrm{eV}]$ & $G_{0} \rho_{F}^{}$ & $\gamma$ [ $10^{-4}$ eV] \\
\hline
LaCe & $0.00$ & $0$ & $0.000$ & $0.01$ & $0.3030$ & 0.00\\
LaCe & $0.55$ & $2$ & $0.080$ & $0.01$ & $0.2900$ & 1.20\\
LaCe & $1.00$ & $4$ & $0.095$ & $0.01$ & $0.2810$ & 1.30\\
LaCe & $1.50$ & $5$ & $0.100$ & $0.01$ & $0.2655$ & 1.50\\
LaCe & $2.00$ & $6$ & $0.105$ & $0.01$ & $0.2520$ & 1.50\\
\hline\hline
ThGd & $0.00$ & $0$ & $0.00$ & $0.01$ & $0.2200$ & 0.00\\
ThGd & $0.10$ & $3$ & $0.11$ & $0.01$ & $0.2090$ & 0.20\\
ThGd & $0.20$ & $5$ & $0.18$ & $0.01$ & $0.1915$ & 0.30\\
\hline
\end{tabular}
\end{center}
\caption{\label{table4} The parameter values exploited during the numerical analysis of Eqs.~(\ref{Delta}), (\ref{nuEq}) and (\ref{SquaredCriticalField}).}
\end{table}

\section{The effect of an external magnetic field on the superconducting alloys}
\label{Sec:SCH}
In the presence of an external magnetic field $\mathcal H$, the system's Hamiltonian (Eq.~(\ref{Hsd})) should be supplemented by the additional terms, describing the interaction with a magnetic field. This yields \cite{Mattis1985}:
\begin{equation}
\label{H_H}
H^{(M)}({\mathcal H}) = H^{(M)} - \frac{1}{2} g^{\prime} \mu_{\mathrm{B}} \tilde{{\mathcal H}} \sum_{\alpha} S_{z\alpha} - \mu_{\mathrm{B}} {\mathcal H} \sigma_{z},
\end{equation}
where $g^{\prime}$ is the modified Land\'{e} factor \cite{WF75}, $\mu_{\mathrm{B}}$ denotes the Bohr magneton, $\tilde{{\mathcal H}}=g_{0}{\mathcal H}$ is the effective magnetic field at each impurity site.

The additional electron $\sigma_{z}$ and impurity spin $S_{z}$ operators in Eq.~(\ref{H_H}) are separated. This allows to use thermodynamic equivalence of Hamiltonians $H^{(M)}$ and $h^{(M)}(\nu, \eta)$, proven in Sec.~\ref{Sec:UpperBound} to describe the thermodynamic properties of a BCS superconductor containing the magnetic impurities and in the presence of magnetic field in terms of the Hamiltonian $h^{(M)}(\mathcal H)$, which has the following form:
\begin{equation}
\label{h_H}
h^{(M)}({\mathcal H}) =  h^{(M)}(\nu, \eta) - \frac{1}{2} g^{\prime} \mu_{\mathrm{B}} \tilde{{\mathcal H}} \sum_{\alpha} S_{z\alpha} - \mu_{\mathrm{B}} {\mathcal H} \sigma_{z}.
\end{equation}
Accordingly, the electrons and impurities are described by the Hamiltonians $\tilde{h}$ and $h_{imp}$ respectively,
\begin{equation}
\begin{split}
\label{h_H2}
\tilde{h} & = H_{\mathrm{BCS}} + (\kappa - \mu_{\mathrm{B}}{\mathcal H}) \sigma_{z} \\ &= H_{\mathrm{BCS}} - \bigl(g M f^{(S)}_{2} + \mu_{\mathrm{B}}{\mathcal H}\bigr) \sigma_{z} \\ &= H_{\mathrm{BCS}} - \lambda \sigma_{z},
\end{split}
\end{equation}

\begin{equation}
\begin{split}
\label{himp_H}
h_{imp}({\mathcal H}) & = g \left(\nu - \frac{1}{2g} g^{\prime}\mu_{\mathrm{B}} \tilde{{\mathcal H}} \right) \sum_{\alpha} S_{z\alpha} + \frac{1}{2}N^{-1} g^{2} \sum_{\alpha} S_{z\alpha}^{2} \\& = g \zeta \sum_{\alpha} S_{z \alpha} + \frac{1}{2} g^{2} N^{-1} \sum_{\alpha} S_{z\alpha}^{2}.
\end{split}
\end{equation}

The form of equations (\ref{h_H}) and (\ref{himp_H}) is very similar to Eqs.~(\ref{hEl}) and (\ref{hImp}). It follows that, in order to include the effect of an external magnetic field on the free energy of a BCS superconductor containing magnetic impurities it suffices to perform the following substitutions in Eq.~(\ref{F_1_2}):
\begin{equation}
\label{H_podstawienie1}
\kappa \rightarrow -\lambda = \kappa - \mu_{\mathrm{B}} {\mathcal H},
\end{equation}
\begin{equation}
\label{H_podstawienie2}
\nu \rightarrow \zeta = \nu - \frac{1}{2g} g^{\prime} \mu_{\mathrm{B}} \tilde{\mathcal H}.
\end{equation}
Accordingly, the set of equations for the parameters $\Delta$ (Eq.~(\ref{Delta})) and $\nu$ (Eq.~(\ref{nuEq})) in the presence of the external magnetic field take the form:
\begin{equation}
\label{DeltaH}
\Delta = \frac{1}{2} G_{0} \rho_{F}^{} \int_{-\delta}^{\delta} \frac{\Delta}{E} f_{3}\bigl(\beta, E, \xi, f^{(S)}_{2}(\zeta), {\mathcal H}\bigr) \mathrm{d} \xi,
\end{equation}
\begin{equation}
\label{nuEqH}
\nu = f_{1}\bigl(\beta, E, \xi, f^{(S)}_{2}(\zeta), {\mathcal H})\bigr) + f^{(S)}_{2}(\zeta, {\mathcal H}), \qquad S = 1/2,\ 7/2,
\end{equation}
where
\begin{equation}
\label{f1H}
f_{1}\bigl(\beta, E, \xi, f^{(S)}_{2}(\zeta), {\mathcal H}\bigr) = \frac{c g}{M} \frac{\sinh \left[ \beta \left(g M f^{(S)}_{2}(\zeta) + \mu_{\mathrm{B}} {\mathcal H}\right) \right]}{\cosh \left[ \beta \left(g M f^{(S)}_{2}(\zeta) + \mu_{\mathrm{B}} {\mathcal H}\right) \right] + \cosh(\beta E)},
\end{equation}
\begin{equation}
\label{f3H}
f_{3}\bigl(\beta, E, \xi, f^{(S)}_{2}(\zeta), {\mathcal H}\bigr) = \frac{\sinh(\beta E)}{\cosh(\beta E) + \cosh \left[ \beta \left(g M f^{(S)}_{2}(\zeta) + \mu_{\mathrm{B}} {\mathcal H}\right) \right]}.
\end{equation}
Functions $f^{(S)}_{2}(\zeta)$ are given by equations (\ref{f2_1_2}), (\ref{f2_7_2}), with $\zeta$ replacing $\nu$.

The free energy of the BCS superconductor perturbed by magnetic impurities and in the presence of an external magnetic field $\mathcal{H}$ then reads
\begin{equation}
\label{F_1_2_H}
\begin{split}
F^{(S)}(\mathcal{H}) = & \min_{\{\Delta,\,\nu\}} \Bigl\{ \rho_{F}^{} |\Lambda| \int_{-\delta}^{\delta} \Bigl[ \frac{1}{2} \Delta^{2} E^{-1} f_{3}\bigl(\beta, E, \xi, f_{2}^{(S)}(\zeta), {\mathcal H}\bigr) - \beta^{-1} \ln \bigl[2 \cosh(\beta E) \\ & + 2 \cosh\left[ \beta \left(g M f^{(S)}_{2}(\zeta) + \mu_{\mathrm{B}} {\mathcal H}\right) \right]\bigr]\Bigr]\mathrm{d}\xi + M^{2}c^{-1} \bigl(\nu f_{2}^{(S)}(\zeta) - \frac{1}{2} \bigl(f_{2}^{(S)}(\zeta)\bigr)^{2}\bigr) \\&+ F_{imp}^{(S)}({\mathcal H}) + E_{0}(\Delta = 0) + \rho_{F}^{} \delta^{2}\Bigr\}, \qquad S = 1/2, 7/2,
\end{split}
\end{equation}
where $F_{imp}^{(S)}({\mathcal H})$ are given by Eqs.~(\ref{Fimp_1_2}) and (\ref{Fimp_7_2}) after substitution $\zeta \rightarrow \nu$.

\section{Critical temperature}
\label{Sec:CriticalTemperature}
The phase diagrams of a BCS superconductor perturbed by magnetic impurities depicted by us in Ref.~\onlinecite{DB11} show that, the phase transition from the normal (Non $SC$) to a supercondcting state can be of the first or second order, depending on the value of the magnetic coupling constant $g$. The next two subsections are concerned with computation of the transition temperature $T_{\mathrm{c}}({\mathcal H})$ for first and second order phase transitions.

\subsection{Second order phase transitions}
According to section~\ref{Sec:MeanFieldDescription}, Eq.~(\ref{Delta}) for the solution $\{\Delta \neq 0, \nu = 0\}$ reduces to the BCS gap equation
\begin{equation}
\label{DeltaBCS}
\Delta_{\mathrm{BCS}} = \frac{1}{2} G_{0} \rho_{\mathrm{F}} \int_{-\delta}^{\delta} \frac{\Delta_{\mathrm{BCS}}}{E_{\mathrm{BCS}}} \tanh \left(\frac{1}{2}\beta E_{\mathrm{BCS}}\right) \mathrm{d}\xi,
\end{equation}
\[
E_{\mathrm{BCS}} = \sqrt{\xi^{2} + \Delta_{\mathrm{BCS}}^{2}}.
\]
The transition temperature $T^{(\mathrm{BCS})}_{\mathrm{c}}$ in BCS theory, is defined as the boundary of the region beyond which there is no real, positive $\Delta_{\mathrm{BCS}}$ satisfying Eq.~(\ref{DeltaBCS}). Below $T^{(\mathrm{BCS})}_{\mathrm{c}}$ the solution $\Delta_{\mathrm{BCS}} \neq 0$ minimizes the free energy and the system is in superconducting phase. Therefore, $T^{(\mathrm{BCS})}_{\mathrm{c}}$ can be obtained from Eq.~(\ref{DeltaBCS}) with $\Delta_{\mathrm{BCS}} = 0$, which yields~\cite{BCS}:
\begin{equation}
\label{TcBCS}
T^{(\mathrm{BCS})}_{\mathrm{c}} = 1.14 \delta \exp \bigl[-(G_{0} \rho_{\mathrm{F}} )^{-1}\bigr].
\end{equation}
It should be possible to estimate the change in $T^{(\mathrm{BCS})}_{\mathrm{c}}$, since the density of states enters exponentially in Eq.~(\ref{TcBCS}). However, significant deviations from Eq.~(\ref{TcBCS}) were observed experimentally for a number of superconductors containing magnetic impurities. This inadequacy of Eq.~(\ref{TcBCS}) is most distinct for large values of impurity concentration. BCS theory is therefore incapable to describe the superconducting alloys.

Expression for transition temperature $T_{\mathrm{c}}$ of a superconducting alloy in the presence of an external magnetic field for 2nd order phase transition can be computed analogously as in BCS theory. To this end, it suffices to put $\Delta = 0$ in Eqs.~(\ref{DeltaH}) and (\ref{nuEqH}). Thus, one obtains the following set of equations for $T_{\mathrm{c}} = 1/(k \beta_{\mathrm{c}}$):
\begin{equation}
\label{DeltaHC}
2 = G_{0} \rho_{F}^{} \int_{-\delta}^{\delta} \frac{\mathrm{d}\xi}{|\xi|} \frac{\sinh(\beta_{\mathrm{c}} |\xi|)}{\cosh(\beta_{\mathrm{c}} |\xi|) + \cosh\left[\beta_{\mathrm{c}} \left( g M f^{(S)}_{2}(\zeta_{\mathrm{c}}) + \mu_{\mathrm{B}} {\mathcal H}\right)\right]},
\end{equation}
\begin{equation}
\label{NuHC}
\nu_{\mathrm{c}} = \frac{c g}{M} \frac{\sinh\left[\beta_{\mathrm{c}} \left( g M f^{(S)}_{2}(\zeta_{\mathrm{c}}) + \mu_{\mathrm{B}} {\mathcal H}\right)\right]}{\cosh\left[\beta_{\mathrm{c}} \left( g M f^{(S)}_{2}(\zeta_{\mathrm{c}}) + \mu_{\mathrm{B}} {\mathcal H}\right)\right] + \cosh(\beta_{\mathrm{c}} |\xi|)} + f^{(S)}_{2}(\zeta_{\mathrm{c}}),
\end{equation}
where $\zeta_{\mathrm{c}} = \nu_{\mathrm{c}} - \frac{1}{2}g^{-1}g^{\prime}\mu_{\mathrm{B}}\tilde{{\mathcal H}}$, $\nu_{\mathrm{c}} = \nu(\beta_{\mathrm{c}})$, $S = 1/2,\,7/2$.

Numerical analysis shows that in the low-temperature scale $\nu_{\mathrm{c}}(T)$ is almost independent in $T$, viz. $\nu_{\mathrm{c}}(T) \approx \nu(0) = c g / M$. Accordingly, the set of equations (\ref{DeltaHC}),~(\ref{NuHC}) is solved under the assumption, that $\nu_{\mathrm{c}} = c g / M$.

\begin{figure}
\begin{center}
\begin{tabular}{@{}c@{ }c@{ }c@{ }c@{}@{ }@{ }c@{ }c@{ }c@{ }c@{ }}
\multicolumn{1}{l}{\footnotesize \bf a} & \multicolumn{1}{l}{\footnotesize{\bf b}}\\[-1.0cm]
    \includegraphics[width=0.5\textwidth]{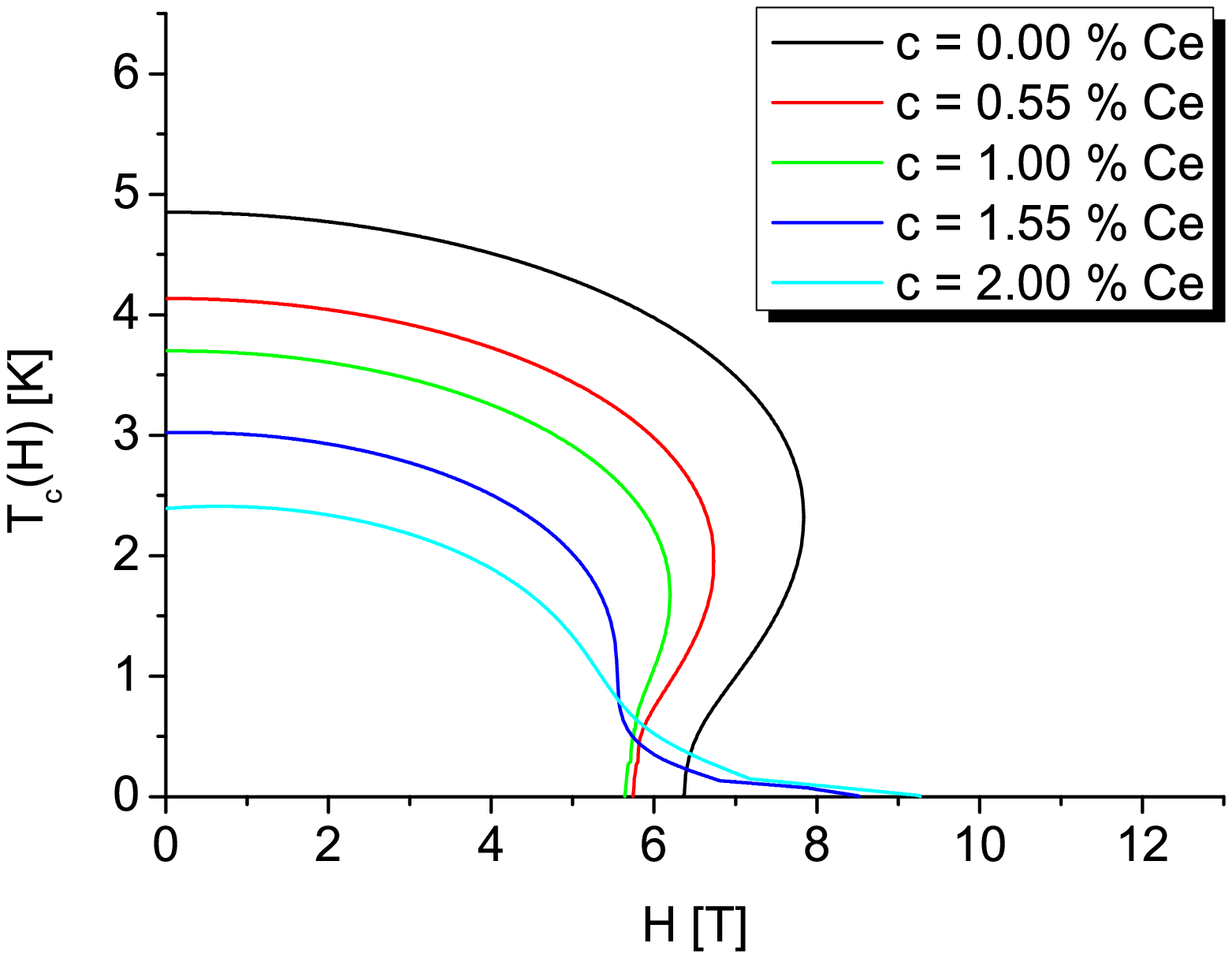}&
    \includegraphics[width=0.5\textwidth]{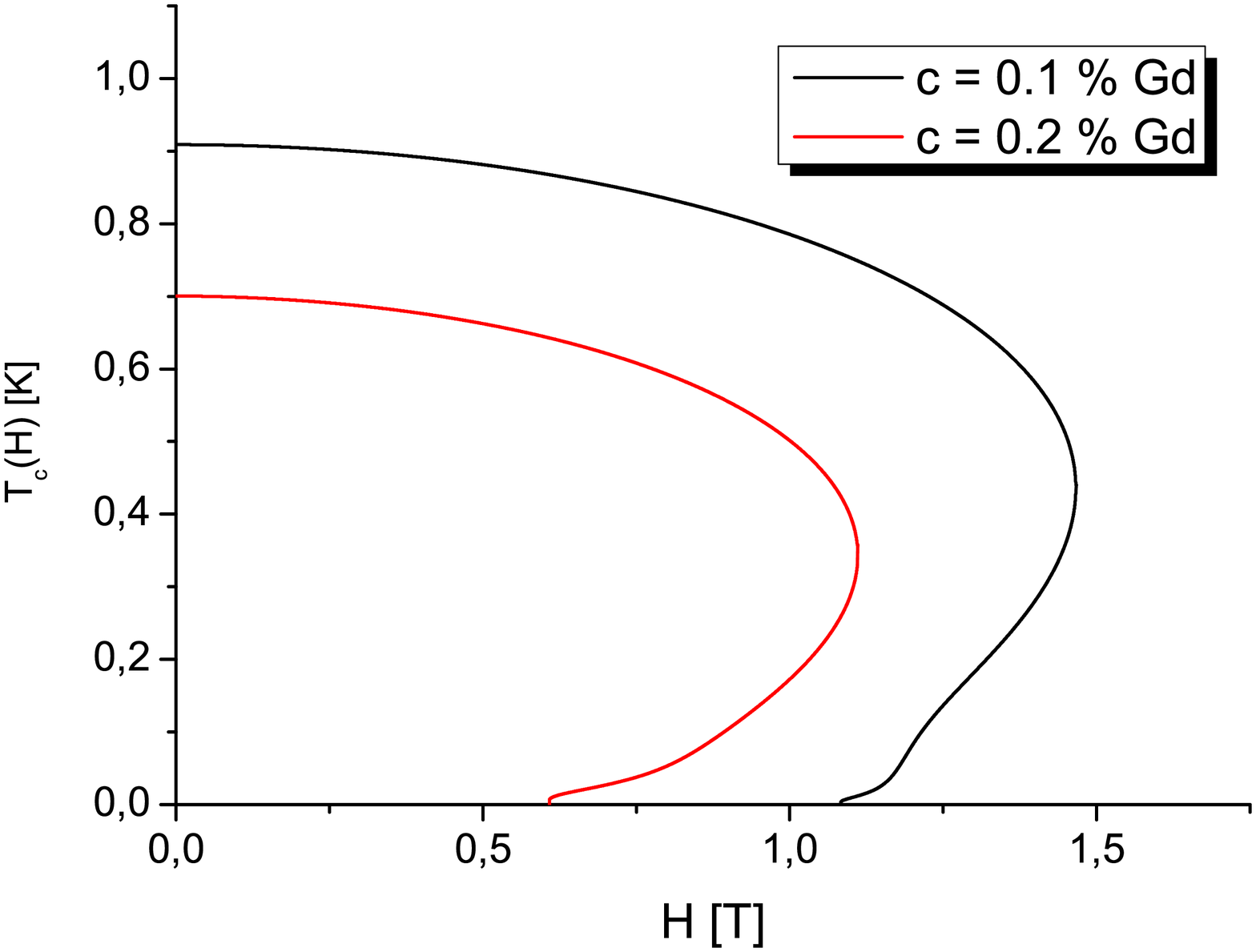}\\[-0.3cm]
\end{tabular}
\caption{\label{F:TcH1}$T_{\mathrm{c}}({\mathcal H})$ graphs under varying impurity concentration for LaCe (a) and ThGd (b). The parameter values $\delta$, $g$, $g_{0}$, $g^{\prime}$, $G_{0}\rho_{F}$ are $M$ collected in Table~\ref{tablePDH}.}
\end{center}
\end{figure}

The resulting solution for $T_{\mathrm{c}}({\mathcal H})$ for $S=1/2, 7/2$ under varying impurity concentration is depicted in Fig.~\ref{F:TcH1}. The solution for $T_{\mathrm{c}}({\mathcal H})$ for small $c$ is similar to the numerical result obtained by Sarma~\cite{Sarma63} of the system described by the Hamiltonian $H_{\mathrm{S}} = H_{\mathrm{BCS}} + \mu_{\mathrm{B}} {\mathcal H} \sigma_{z}$. His result for $T^{(BCS)}_{\mathrm{c}}({\mathcal H})$ agrees qualitatively with $T_{\mathrm{c}}({\mathcal H})$ graphs depicted in Fig.~\ref{F:TcH1}, since the expression for $T^{(BCS)}_{\mathrm{c}}({\mathcal H})$ obtained in Ref.~\onlinecite{Sarma63} is of the similar form to Eq.~(\ref{DeltaHC}) with $\zeta_{\mathrm{c}} = 0$.

The $T_{\mathrm{c}}({\mathcal H})$ graphs depicted in Fig.~\ref{F:TcH1}a show two phase transitions for sufficiently small impurity concentration, i.e. $c \in (0\,\mathrm{at.}\,\%, 0.10\,\mathrm{at.}\,\%)$ and for ${\mathcal H} \in (5.65 \,\mathrm{T}, 7.8 \,\mathrm{T})$). The first phase transition ($P \rightarrow SC$) occurs at $T_{\mathrm{c}1}({\mathcal H})$ and the second phase transition (back to normal state) appears at $T_{\mathrm{c}2}({\mathcal H})$. In case of LaCe, only one phase transition is present ($SC \rightarrow P$) for higher concentrations, but superconductor overcomes larger values of an external magnetic field. Furthermore,  $T_{\mathrm{c}}({\mathcal H})$ initially increases with ${\mathcal H}$.

The form of denominator on the right hand side of Eq.~(\ref{DeltaHC}) suggests that, the perturbative effect of magnetic impurities can be compensated by an external magnetic field. It follows from the fact that $f_{2}^{(S)}(\zeta_{\mathrm{c}})$ is odd function in $\zeta_{\mathrm{c}}$ and from a definition of $\zeta_{\mathrm{c}}$ parameter, which approaches negative values for sufficiently large ${\mathcal H}$. Thus, the magnetic field intensity, required for the full compensation of the perturbative effect of magnetic imurities on a BCS supercondutor has the form:
\begin{equation}
{\mathcal H}_{\mathrm{k}} = - \frac{g M}{\mu_{\mathrm{B}}} f_{2}^{(S)}(\zeta_{\mathrm{c}}).
\end{equation}

\begin{figure}
\begin{center}
\begin{tabular}{@{}c@{ }c@{ }c@{ }c@{}@{ }@{ }c@{ }c@{ }c@{ }c@{ }}
\multicolumn{1}{l}{\footnotesize \bf a} & \multicolumn{1}{l}{\footnotesize{\bf b}}\\[-1.0cm]
    \includegraphics[width=0.5\textwidth]{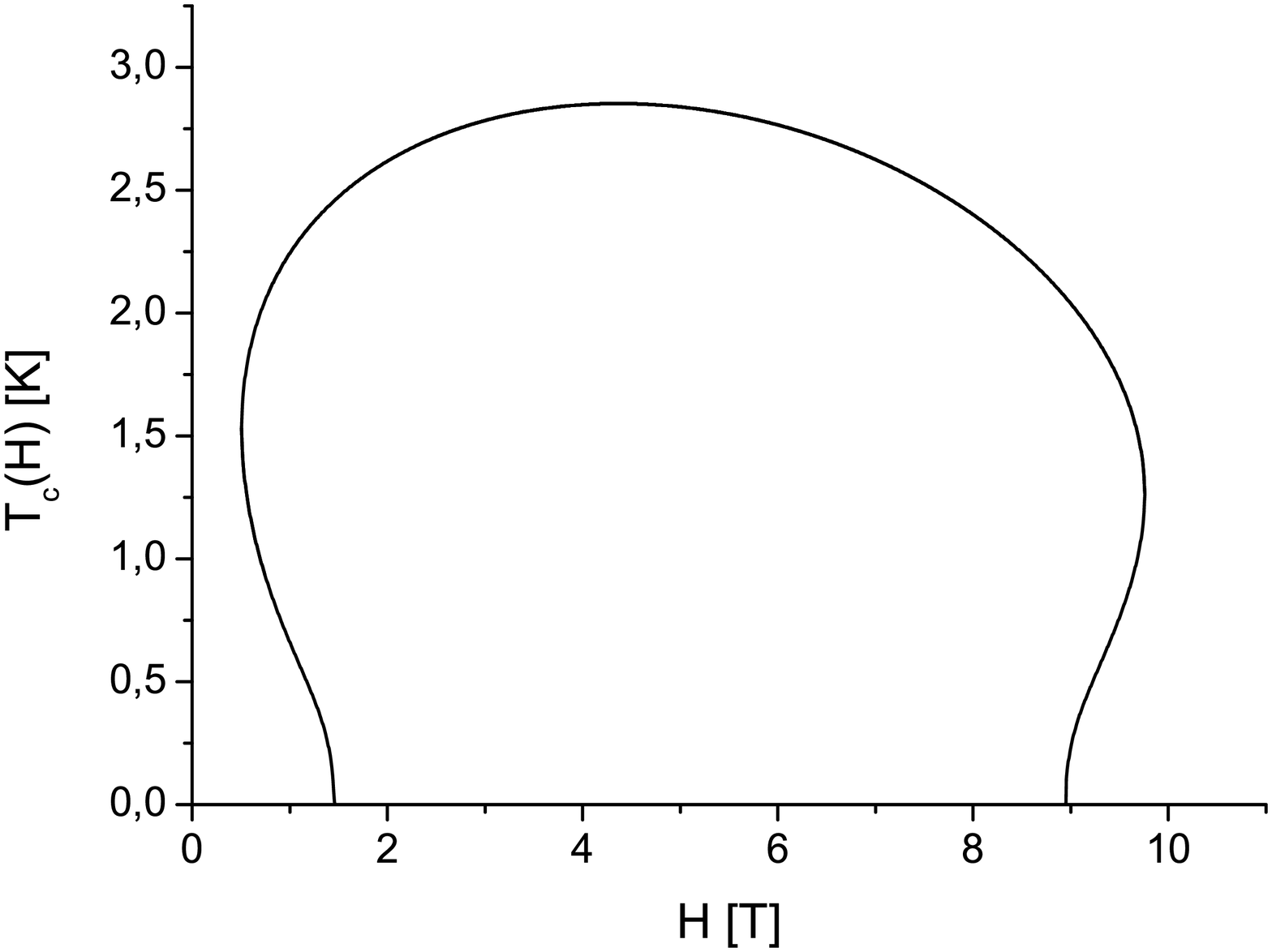}&
    \includegraphics[width=0.5\textwidth]{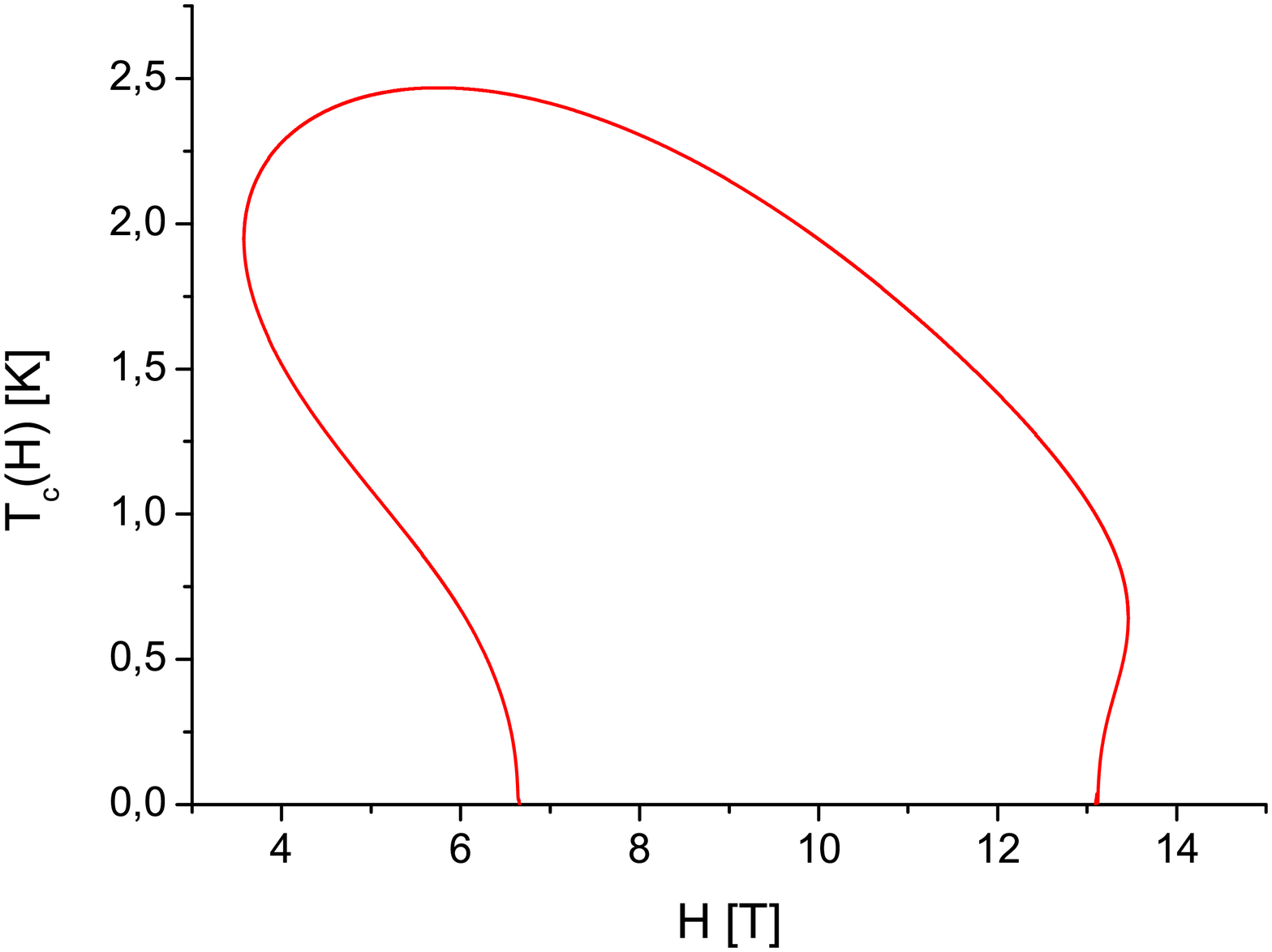}\\
\multicolumn{1}{l}{\footnotesize \bf c} & \multicolumn{1}{l}{\footnotesize{\bf d}}\\[-1.0cm]
    \includegraphics[width=0.5\textwidth]{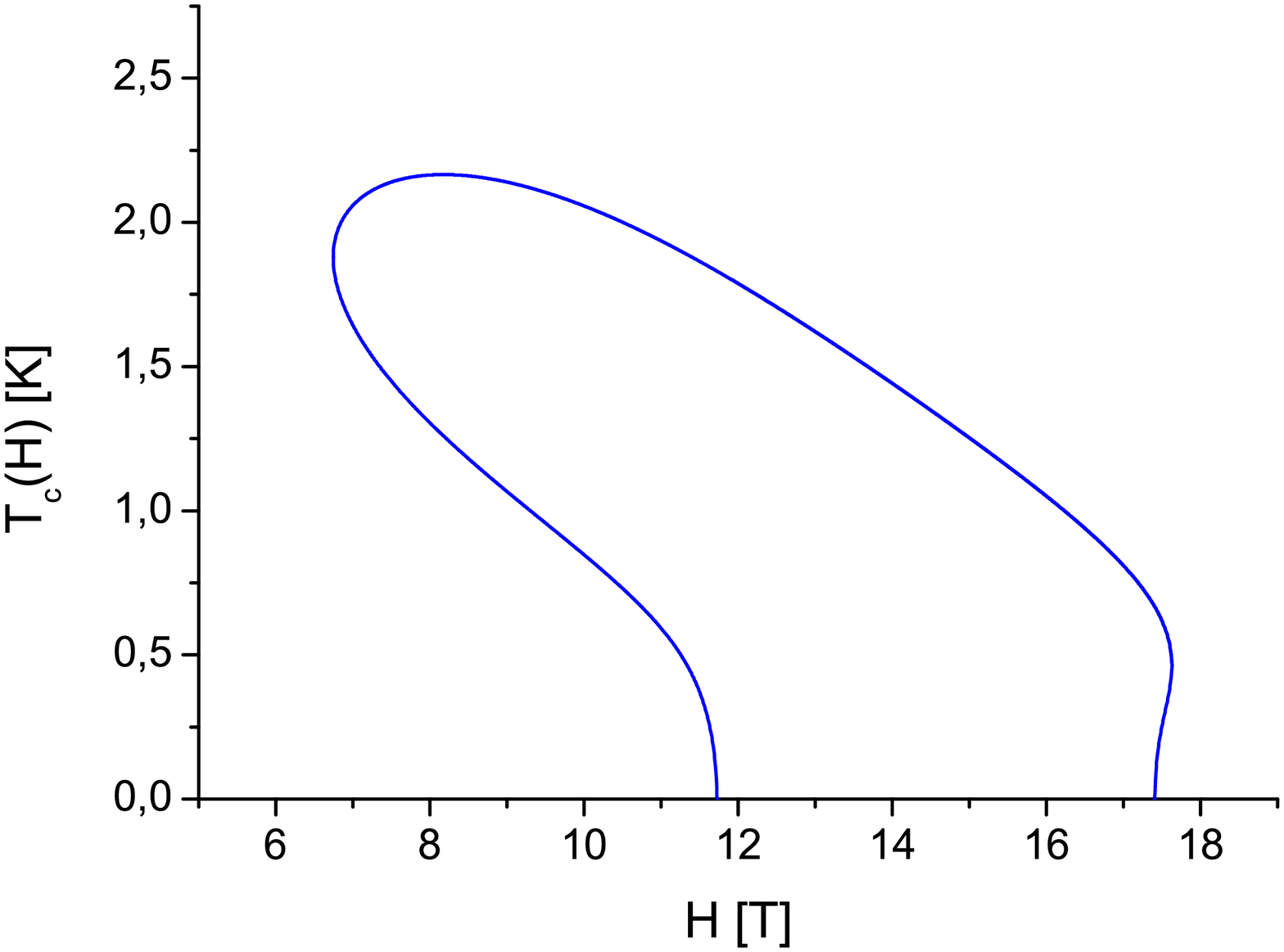}&
    \includegraphics[width=0.5\textwidth]{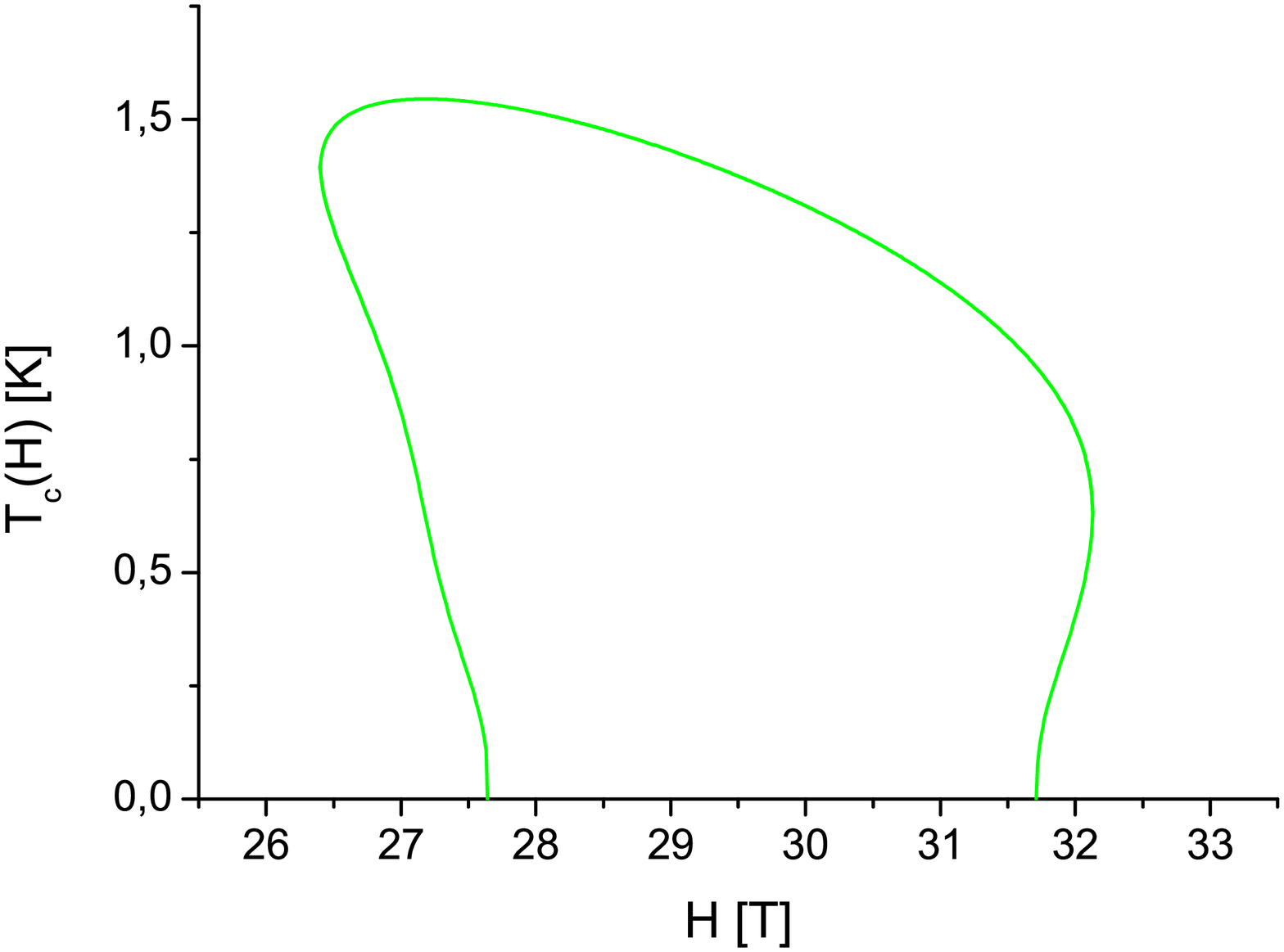}\\[-0.3cm]
\end{tabular}
\caption{\label{F:TcH2}$T_{\mathrm{c}}({\mathcal H})$ graphs of (La$_{1-x}$Ce$_{x}$)Al$_{2}$ for $g=0.95\,\sqrt{\mathrm{eV}}$ and under varying impurity concentration: (a) $x = 0.0010$, (b) $x = 0.0019$, (c) $x = 0.0028$, (d) $x = 0.0057$. The values of the parameters $\delta$, $g_{0}$, $g^{\prime}$, $G_{0}\rho_{F}$ and $M$ are collected in Table~\ref{tablePDH}}
\end{center}
\end{figure}

This supposition has been verified for $g = 0.95 \,\sqrt{\mathrm{eV}}$ and various impurity concentrations. The values of remaining parameters correspond to (La$_{1-x}$Ce$_{x}$)Al$_{2}$ alloy (Table~\ref{tablePDH}). The results, which are depicted in Fig.~\ref{F:TcH2}, confirm the hypothesis of the Jaccarino-Peter compensation effect in the considered theoretical model. According to Fig.~\ref{F:TcH2}, the values of $T_{\mathrm{c}1}({\mathcal H})$ increases with $\mathcal H$ and after reaching a maximum at ${\mathcal H} = {\mathcal H}_{\mathrm{k}}$, decreases and finally falls to zero.

The value of an exchange field ${\mathcal H}_{\mathrm{J}}$ is expected to increase with increasing impurity concentration, since the number of the magnetic moments of impurities, which are antiferromagnetically coupled to conduction fermions also increases. As a result the value of an external magnetic field ${\mathcal H}$ required to compensat the perturbative effect of ${\mathcal H}_{\mathrm{J}}$ increases with $c$.

\begin{table}
\tabcolsep8pt
\begin{center}
\begin{tabular}{*{9}{|c}|}
\hline
Alloy & $x$ & c [\%]& $M$ & $\delta\ [\mathrm{eV}]$ & $G_{0} \rho_{F}^{}$ & $g^{\prime}$ & $g_{0}$ & $g$ [$\sqrt{\mathrm{eV}}$]\\
\hline\hline
\multirow{5}{*}{LaCe} & \multirow{5}{*}{--} & $0.00$ & $0$ & \multirow{5}{*}{$0.01$} & $0.3030$ & \multirow{5}{*}{$10/7$} & \multirow{5}{*}{$0.006$} & $0.00$ \\
& & $0.55$ & $2$ & & $0.2890$ & & & $0.080$ \\
& & $1.00$ & $4$ & & $0.2800$ & & & $0.095$ \\
& & $1.55$ & $5$ & & $0.2650$ & & & $0.100$ \\
& & $2.00$ & $6$ & & $0.2500$ & & & $0.105$ \\
\hline\hline
\multirow{4}{*}{(La$_{1-x}$Ce$_{x})$Al$_{2}$} & $0.0010$ & \multirow{4}{*}{--} & $1$ & \multirow{4}{*}{$0.01$} & $0.2610$ & \multirow{4}{*}{$10/7$} & \multirow{4}{*}{$0.006$} & $0.10$ \\
& $0.0019$ & & $4$ & & $0.2515$ & & & $0.189$ \\
& $0.0028$ & & $7$ & & $0.2435$ & & & $0.19$ \\
& $0.0057$ & & $8$ & & $0.2250$ & & & $0.23$ \\
\hline \hline
\multirow{3}{*}{ThGd} & \multirow{3}{*}{--} & $0.0$ & $0$ & \multirow{3}{*}{$0.01$} & $0.2200$ & \multirow{3}{*}{$2$} & \multirow{3}{*}{$0.004$} & $0.00$ \\
 & & 0.1 & $3$ & & $0.2010$ & & & $0.11$ \\
 & & 0.2 & $5$ & & $0.1915$ & & & $0.18$ \\
\hline
\end{tabular}
\caption{\label{tablePDH}The parameter values.}
\end{center}
\end{table}

The $T_{\mathrm{c}}({\mathcal H})$ graph depicted in Fig.~\ref{F:TcH2}a resembles the dependence of superconducting transition temperature $T_{\mathrm{c}}$ on dopant concentration $p$, exhibited by high-temperature cuprates, e.g. $\mathrm{La}_{2-x}\mathrm{Sr}_x\mathrm{CuO}_4$ \cite{MD89, YL98}, $\mathrm{YBa}_2\mathrm{Cu}_3\mathrm{O}_y$ \cite{UE91}, $\mathrm{Bi}_{2-x}\mathrm{Pb}_x\mathrm{Sr}_2\mathrm{Ca}_2\mathrm{Cu}_3\mathrm{O}_{10}$~\cite{UE91},
$\mathrm{Bi}_2\mathrm{Sr}_{2-x}\mathrm{La}_x\mathrm{CuO}_6$ \cite{AM99} and iron-pnictides~\cite{Hosono}. $T_{\mathrm{c}}(p)$ initially increases almost linearly in $p$ and after reaching a maximum at optimal doping level $p_{opt}$, decreases and finally falls to zero.

Analogous behavior of the superconducting transition temperature is observed, e.g. in CeRhIn$_{5}$ \cite{YK02}, CeCoIn$_{5}$ \cite{VS02}, when the superconductor is under pressure.

\subsection{First order phase transitions}
In the case, of first order phase transitions, the assumption that the gap parameter $\Delta$ vanishes at the transition temperature does not hold. According to the results obtained in Ref.~\onlinecite{DB11}, the superconducting transition temperature $T_{\mathrm{c}}$ possesses three solutions ($T_{\mathrm{c}1} \geq T_{\mathrm{c}2} \geq T_{\mathrm{c}3}$) for certain values of $g$ and $c$. These solutions can be determined numerically from the following equations:
\begin{equation}
\label{Tc_1_1szyRodzajH}
T_{\mathrm{c}1} : \qquad F_{P}({\mathcal H}) - F_{SC}({\mathcal H}) = 0,
\end{equation}
\begin{equation}
\label{Tc_2_1szyRodzajH}
T_{\mathrm{c}2} : \qquad F_{SC}({\mathcal H}) - F_{\Phi}({\mathcal H}) = 0, \qquad \Phi = D, F,
\end{equation}
\begin{equation}
\label{Tc_3_1szyRodzajH}
T_{\mathrm{c}3} : \qquad F_{F}({\mathcal H}) - F_{D}({\mathcal H}) = 0.
\end{equation}

The existence of $T_{\mathrm{c}3}$ depends on the type of phase transition occuring at $T_{\mathrm{c}2}$. If~the system undergoes a phase transition to ferromagnetic phase at $T_{\mathrm{c}2}$, then $T_{\mathrm{c}3} > 0$ for certain values of $g$. If $T_{\mathrm{c}2} = T_{SCD}$, then $T_{\mathrm{c}3} = 0$ and the system does not reenter the superconducting phase ($SC$ or $D$).

\begin{figure}
\begin{center}
\includegraphics[width=0.7\textwidth]{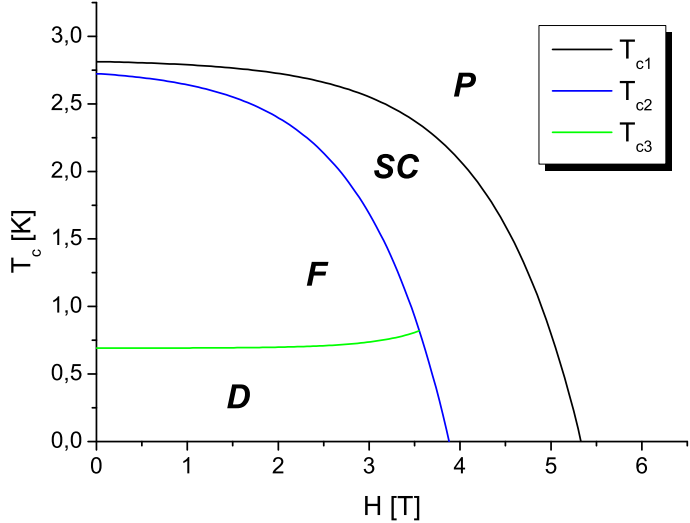}
\caption{\label{F:TcPierwszyRodzajH1}$T_{\mathrm{c}}({\mathcal H})$ graphs of (La$_{1-x}$Ce$_{x}$)Al$_{2}$ for $g=0.5\,\sqrt{\mathrm{eV}}$ and $x = 0.0010$. The values of the parameters $\delta$, $g_{0}$, $g^{\prime}$, $G_{0}\rho_{F}$ and $M$ are collected in Table~\ref{tablePDH}. For ${\mathcal H} \in (0\,\mathrm{T}, 3.55\,\mathrm{T})$ the system undergoes three phase transitions with decreasing temperature, $P \rightarrow SC \rightarrow F \rightarrow D$, showing the Jaccarino-Peter compensation effect.}
\end{center}
\end{figure}

The solution of Eqs.~(\ref{Tc_1_1szyRodzajH})--(\ref{Tc_3_1szyRodzajH}) for the parameter values corresponding to (La$_{1-x}$Ce$_{x}$)Al$_{2}$ \cite{DB11} are depicted in Fig.~\ref{F:TcPierwszyRodzajH1}. For ${\mathcal H} \in (0\,\mathrm{T}, \, 3.55 \mathrm{T})$, three phase transitions $P \rightarrow SC \rightarrow F \rightarrow D$ with decreasing temperature are present. These phase transitions can be interpreted as $\mathrm{Non}\,SC \rightarrow SC \rightarrow \mathrm{Non}\,SC \rightarrow SC$ transitions, i.e. the Jaccarino-Peter compensation effect, which has been experimentally observed in a number of superconducting magnetic alloys, e.g.: Sn$_{x}$Eu$_{1.2-x}$Mo$_{6}$S$_{8}$ \cite{SW82}, Eu$_{0.75}$Sn$_{0.25}$Mo$_{6}$S$_{7.2}$Se$_{0.8}$ \cite{HM84}, CeCoIn$_{5}$ \cite{HR03}, URhGe i UCoGe \cite{DA011}.

\section{Concluding remarks}
Our recent work on the phase diagrams of a BCS superconductor perturbed by a reduced s-d interaction \cite{DB11} was continued to examine the critical magnetic field and the effect of an external magnetic field ${\mathcal H}$ on the superconducting transition temperature $T_{\mathrm{c}}$ of such system. The superconductivity enhancement, revealed by an increase of the critical magnetic field with decreasing temperature and increasing impurity concentration, has been found.

Good quantitative agreement of the resulting theoretical expressions for the critical magnetic field with experimental data was demonstrated for LaCe, ThGd and SmRh$_{4}$B$_{4}$.

The numerical analysis of $T_{\mathrm{c}}({\mathcal H})$ showed that the perturbative effect of magnetic impurities can be compensated by an external magnetic field, providing the evidence of the Jaccarino-Peter effect, which has been experimentally observed in a number of superconducting magnetic alloys, e.g.: Sn$_{x}$Eu$_{1.2-x}$Mo$_{6}$S$_{8}$ \cite{SW82}, Eu$_{0.75}$Sn$_{0.25}$Mo$_{6}$S$_{7.2}$Se$_{0.8}$ \cite{HM84}. The enhancement of superconductivity, displayed by an increase of the the upper critical magnetic field with decreasing temperature due to the Jaccarino-Peter effect was discovered in SmRh$_{4}$B$_{4}$ \cite{HWMFM78}, CeCoIn$_{5}$ \cite{HR03}, URhGe and UCoGe \cite{DA011}.

The enhancement of the critical magnetic field $H_{\mathrm{c}}$ in magnetic superconductors, which is due to the interplay of the superconductivity and magnetic order, opens the door to their possible industrial applications. Since the upper critical magnetic field at absolute temperature of PbMo$_{6}$S$_{8}$ magnetic superconductor, i.e. $H_{\mathrm{c}}(0) \approx 60$T \cite{OF82} doubly exceeds the corresponding value of the high-performance, low-temperature superconducting material Nb$_{3}$Sn, which is currently widely-exploited in power applications, e.g. in accelerators available at Fermilab, Brookhaven, DESY and CERN \cite{DL01}.

\section*{Acknowledgements}
Author is grateful the Institute for Theoretical and Applied Physics for supporting his stay during the Eurasia-Pacific Summer School \& Workshop on Strongly Correlated Electrons (Turunc, Turkey, 6-17 August 2012), where part of this research was performed and presented. This work was partially supported by the NCU grant for young researchers no.~1134-F.

\section*{References}

\bibliography{2fermion_2}

\end{document}